\documentclass[conference]{IEEEtran}
\IEEEoverridecommandlockouts
\usepackage{cite}
\usepackage{amsmath,amssymb,amsfonts}
\usepackage{algorithmic}
\usepackage{graphicx}
\usepackage{textcomp}
\usepackage{xcolor}
\usepackage{colortbl}
\usepackage{multirow}
\usepackage{listings}
\usepackage{float}
\usepackage{enumitem}
\usepackage{balance}
\usepackage{framed}
\usepackage{comment}
\usepackage{caption}
\usepackage{subcaption}
\usepackage{listings}
\definecolor{codepurple}{rgb}{0.58,0,0.82}
\definecolor{backcolour}{RGB}{239, 239, 239}
\definecolor{codeorange}{RGB}{191, 94, 45}
\definecolor{codegreenl}{rgb}{0,0.6,0}
\definecolor{codegreend}{RGB}{101, 139, 111}
\definecolor{codegray}{rgb}{0.5,0.5,0.5}

\lstdefinestyle{mystyle}{
    backgroundcolor=\color{backcolour},   
    commentstyle=\color{codepurple},
    keywordstyle=\color{codeorange},
    numberstyle=\small\color{codegray},
    stringstyle=\color{codegreenl},
    basicstyle=\fontsize{8}{8}\selectfont\ttfamily,
    breakatwhitespace=false,     
    breaklines=true,                 
    captionpos=b,                    
    keepspaces=true,                 
    numbers=none,                    
    numbersep=2pt,        
    numbers=left,
    stepnumber=1,
    showspaces=false,                
    showstringspaces=false,
    showtabs=false,                  
    tabsize=2
}
\def\BibTeX{{\rm B\kern-.05em{\sc i\kern-.025em b}\kern-.08em
    T\kern-.1667em\lower.7ex\hbox{E}\kern-.125emX}}

\begin{document}

\title{Reproducible Performance Optimization of Complex Applications on the Edge-to-Cloud Continuum}

% \author{\IEEEauthorblockN{1\textsuperscript{st} Given Name Surname}
% \IEEEauthorblockA{\textit{dept. name of organization (of Aff.)} \\
% \textit{name of organization (of Aff.)}\\
% City, Country \\
% email address or ORCID}
% \and
% \IEEEauthorblockN{2\textsuperscript{nd} Given Name Surname}
% \IEEEauthorblockA{\textit{dept. name of organization (of Aff.)} \\
% \textit{name of organization (of Aff.)}\\
% City, Country \\
% email address or ORCID}
% \and
% \IEEEauthorblockN{3\textsuperscript{rd} Given Name Surname}
% \IEEEauthorblockA{\textit{dept. name of organization (of Aff.)} \\
% \textit{name of organization (of Aff.)}\\
% City, Country \\
% email address or ORCID}
% }
%\linespread{0.85}https://www.overleaf.com/project/5e85ab6855a8cd0001e28902

% \author{\IEEEauthorblockN{
% Daniel Rosendo\IEEEauthorrefmark{1},
% Pedro Silva\IEEEauthorrefmark{2},
% Matthieu Simonin\IEEEauthorrefmark{1},
% Alexandru Costan\IEEEauthorrefmark{1},
% Gabriel Antoniu\IEEEauthorrefmark{1}}
% \IEEEauthorblockA{\IEEEauthorrefmark{1}University of Rennes, Inria, CNRS, IRISA - Rennes, France\\
% \{daniel.rosendo, matthieu.simonin, alexandru.costan, gabriel.antoniu\}@inria.fr}
% \IEEEauthorblockA{\IEEEauthorrefmark{2}Hasso-Plattner Institut, University of Potsdam - Berlin, Germany, pedro.silva@hpi.de}\\
% }

\author{\IEEEauthorblockN{
Daniel Rosendo\IEEEauthorrefmark{1},
Alexandru Costan\IEEEauthorrefmark{1},
Gabriel Antoniu\IEEEauthorrefmark{1},
Matthieu Simonin\IEEEauthorrefmark{1},\\
Jean‐Christophe Lombardo\IEEEauthorrefmark{2},
Alexis Joly\IEEEauthorrefmark{2},
Patrick Valduriez\IEEEauthorrefmark{2}}
\IEEEauthorblockA{\IEEEauthorrefmark{1}University of Rennes, Inria, CNRS, IRISA - Rennes, France\\
\{daniel.rosendo, alexandru.costan, gabriel.antoniu, matthieu.simonin\}@inria.fr}
\IEEEauthorblockA{\IEEEauthorrefmark{2}University of Montpellier, Inria, CNRS, LIRMM - Montpellier, France\\
\{jean‐christophe.lombardo, alexis.joly, patrick.valduriez\}@inria.fr}\\
}

% \IEEEoverridecommandlockouts
% \IEEEpubid{\makebox[\columnwidth]{978-1-7281-6677-3/20/\$31.00~\copyright2020 European Union\hfill} \hspace{\columnsep}\makebox[\columnwidth]{ }}

\maketitle

% \IEEEpubidadjcol

\begin{abstract}
In more and more application areas, we are witnessing the emergence of complex workflows that combine computing, analytics and learning. They often require a hybrid execution infrastructure with IoT devices interconnected to cloud/HPC systems (aka \emph{Computing Continuum}). Such workflows are subject to complex constraints and requirements in terms of performance, resource usage, energy consumption and financial costs. This makes it challenging to optimize their configuration and deployment.

We propose a methodology to support the optimization of real-life applications on the Edge-to-Cloud Continuum. We implement it as an extension of \textbf{E2C}\textit{lab}, a previously proposed framework supporting the complete experimental cycle across the Edge-to-Cloud Continuum. Our approach relies on a rigorous analysis of possible configurations in a controlled testbed environment to understand their behaviour and related performance trade-offs. We illustrate our methodology by optimizing Pl@ntNet, a world-wide plant identification application. Our methodology can be generalized to other applications in the Edge-to-Cloud Continuum.

\end{abstract}


\section{Introduction}
\label{sec:introduction}
% Describe the open research questions to which we propose an approach.

%Computing Continuum
The continuous increase of IoT devices and captured data requires rethinking where to process data.
%the notion of \emph{where} to process data. 
%For example, \emph{where} to offload tasks or place data from Edge devices to Fog nodes and Cloud/HPC centres in applications like smart surveillance, autonomous vehicles, \emph{etc.}.
Instead of the traditional data center compute model,
%PV: not sure what this means (big data community) and what does it refer to. 
%the Big Data community is increasingly leveraging
one main approach used in big data is to leverage
compute resources distributed at multiple processing points in the system – from endpoint devices at the edge of the network to data centers or HPC systems at its core. This distributed infrastructure, referred to as the \emph{Computing Continuum}~\cite{beckman2020} (or \emph{Digital Continuum}% or the \emph{Transcontinuum}
), combines heterogeneous computing resources that generate and process data across geographically distributed Edge, Fog, and Cloud/HPC infrastructures.

%Apps on the CC
Real-world applications deployed on such hybrid infrastructure (\emph{e.g.,} smart factory~\cite{wang2018adaptive}, autonomous vehicles~\cite{midya2018multi}, among others) typically need to comply with many constraints related to resource consumption (\emph{e.g.,} GPU, CPU, memory, storage and bandwidth capacities), software components composing the application and requirements such as QoS, security, and privacy~\cite{xia2018combining}. Furthermore, optimizing application workflows on distributed and heterogeneous resources (\emph{i.e.,} minimizing processing latency, energy consumption, financial costs, \emph{etc.}) is challenging. The parameter settings of the applications and the underlying infrastructure result in 
%a myriad of configuration possibilities and, consequently, in 
a complex multi-infrastructure configuration search space~\cite{ranjan2018next}.

% \begin{todo}[Alex]
% The concept of "optimality" in this context (which seems central to the paper) needs to be better defined and motivated (especially considering the EuroPar community, very sensitive to these aspects). What is an optimal mapping / configuration? With respect to which metric(s)? 
% \end{todo}

%PV: this sentence is a bit confusing, as it introduces the solution (analysis in a controlled testbed environment) with the word "require", but who says so? us? others? and the problem and existing solutions are explained later. I think the solution should come AFTER.
%Complex configurations
The intricacies of these configurations require,
prior to production-level deployment, analysis in a controlled testbed environment in order to understand their performance trade-offs (\emph{i.e.,} latency and energy consumption, throughput and resource usage, cost and service availability, \emph{etc.})~\cite{bellendorf2020classification, xu2018survey}.

\begin{figure}[t]
  \centering
  \includegraphics[width=0.9\linewidth]{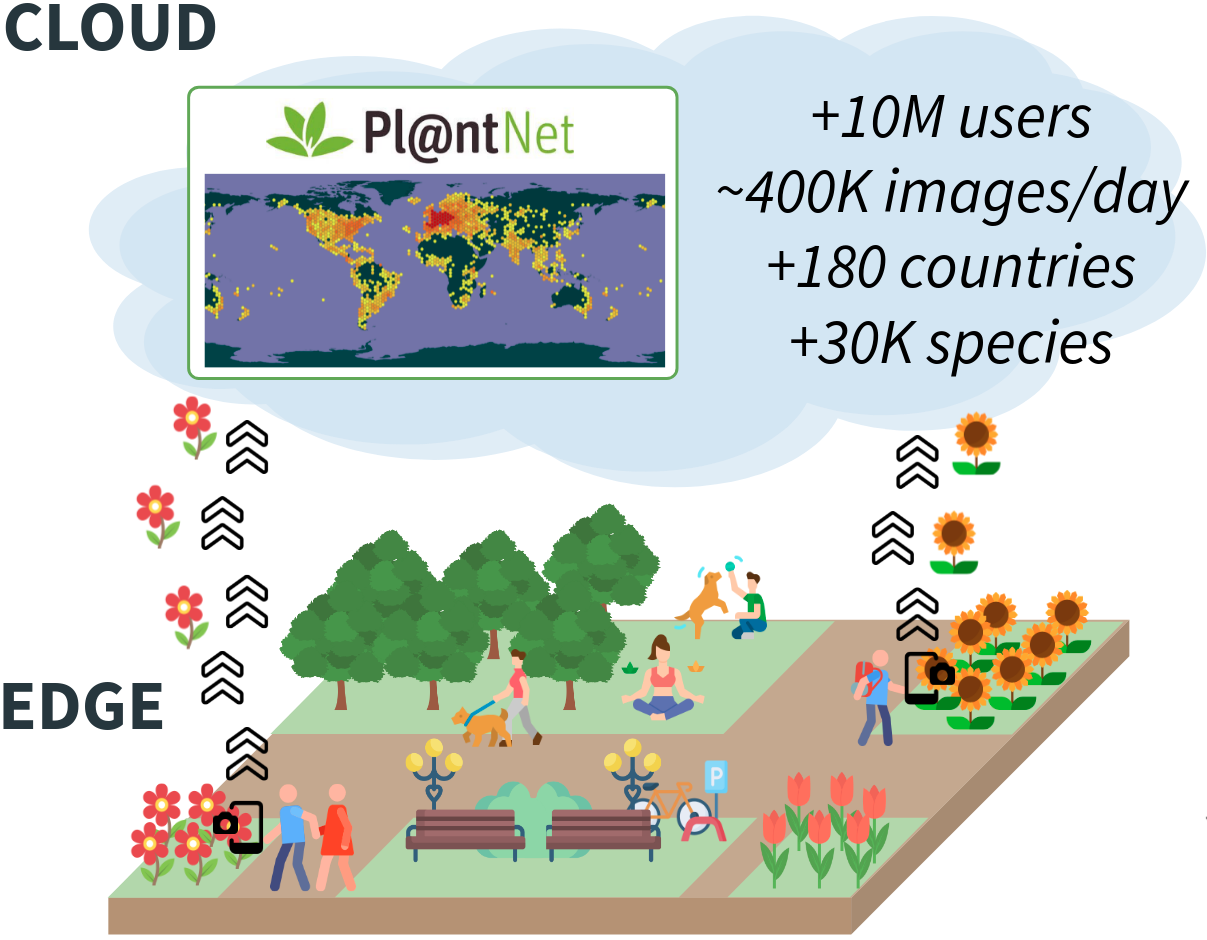}
  \caption{The Pl@ntNet application.}
  \label{fig:plantnet-geographic-scope}
\end{figure}

\begin{figure}[t]
  \centering
  \includegraphics[width=.77\linewidth]{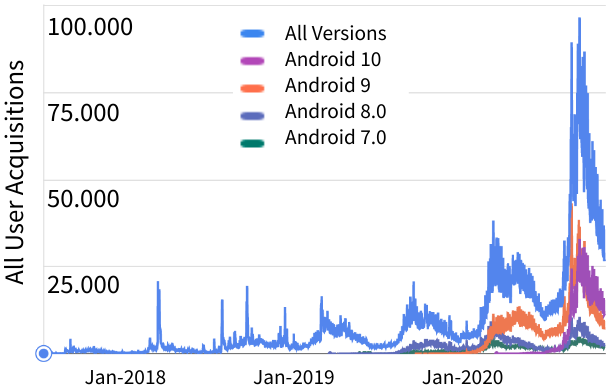}
  \caption{Exponential growth of new users every spring (peaks in May-June).}
  \label{fig:plantnet_new_users}
\end{figure}

Let us illustrate this problem with Pl@ntNet~\cite{joly2016look}, a 
large-scale participatory application for botanical data and AI-based plant identification. Pl@ntNet's main feature is a mobile app that allows smartphone users to identify plants from photos and share their observations (Figure~\ref{fig:plantnet-geographic-scope}). It has more than 10 million users all around the world and processes about 400K plant images per day. One main challenge faced by Pl@ntNet engineers is to anticipate the necessary evolution of the infrastructure to pass the upcoming spring peak (Figure~\ref{fig:plantnet_new_users}) and adapt
%to know what should be done the following years. 
the system configuration to some expected evolution of application usage (\emph{e.g.}, an increase of its number of users). 

There are simulation and emulation tools for the Cloud, Fog, Edge, Fog-to-Cloud, Edge-to-Fog~\cite{calheiros2011cloudsim, gupta2017ifogsim, sonmez2018edgecloudsim, mayer2017emufog}.
However, there is no solution for large-scale deployment and evaluation of real-life applications on testbeds that cover the entire Computing Continuum as a whole and guide application optimization (\emph{i.e.,} minimizing costs, latency, resource consumption, among others) of the entire application workflow.
% \begin{todo}
% [Alex] The last statement seems a bit bold. I think there are some tools/systems/algorithms for "helping" deployments on (at least) hybrid infrastructures Edge/Cloud, Edge/Fog, Fog/Cloud etc. (e.g., Planner). Second, we need to be consistent wrt terminology: configuration optimization that was mentioned earlier != workflow optimization (which is a standalone research domain)
% \end{todo}
%PV: I too have a problem with the last paragraph, which does not follow well the Plantnet paragraph. What I suggest is that, based on Plantnet, you first make clear what is needed, i.e., solution that allows large-scale deployment and evaluation of real-life applications on testbeds. Then, say there is no solution, and mention briefly existing tools.

%PV: can we say something about how hard is the problem? sounds more like engineering.

In this paper, we propose a methodology to support the optimization of real-life applications on the Edge-to-Cloud Continuum. This methodology is useful to help decide on application configurations to optimize relevant metrics (\emph{e.g.,} performance, resource usage, energy consumption, \emph{etc.}) by means of computationally tractable optimization techniques~\cite{pham2012intelligent}.
%PV: not a very strong aim to answer questions
%Our approach aims to answer questions like:
%PV: I propose:
It eases the configuration of the system components distributed on Edge, Fog, and Cloud infrastructures as well as the decision where to execute the application workflow components to minimize communication costs and end-to-end latency.

%\begin{itemize}
%    \item \emph{How to configure the system components distributed on Edge, Fog, and Cloud infrastructures to minimize the processing latency?}
%    \item \emph{Where should the workflow components be executed to minimize communication costs and end-to-end latency?}
    % \item \emph{What is the optimal Fog gateway hardware configuration and Cloud data processing engine software configuration to minimize the end-to-end latency of my application?}
    % \item \emph{What is the optimal number of Fog nodes and Edge device hardware configuration to maximize the overall application performance?}
% \end{itemize}

%We integrated this approach within the \textbf{E2C}\textit{lab} methodology and framework for automatic application deployment
%and reproducible experimentation ~\cite{rosendo2020e2clab}. This paper has the following main contributions:
We implemented this methodology as an extension of the \textbf{E2C}\textit{lab}~\cite{rosendo2020e2clab} framework for automatic application deployment
and reproducible experimentation. This paper has the following main contributions:

\begin{enumerate}
    \item \textbf{A
    %PV: redundant
    %optimization
    methodology to optimize the performance of real-life applications on the Computing Continuum}, leveraging computationally tractable optimization techniques (Section~\ref{sec:methodology}).
    
    \item \textbf{An implementation of this optimization methodology} as an extension of the \textbf{E2C}\textit{lab} framework for reproducible analysis of applications on the Edge-to-Cloud continuum. For this purpose, we enhanced \textbf{E2C}\textit{lab} with an optimization layer. To the best of our knowledge, this enhanced version of \textbf{E2C}\textit{lab} is the first framework to support the complete deployment and analysis cycle of a complex workflow executed on the Computing Continuum (Section~\ref{sec:methodology_impl}).

    \item A \textbf{large scale experimental validation} of the proposed approach with the Pl@ntNet application on 42 nodes of the Grid'5000 testbed~\cite{bolze2006grid}. Our approach helps optimizing Pl@ntNet software configurations across the continuum to minimize user response time (Section~\ref{sec:evaluation}).% The high-level goal of this analysis is to allow Pl@ntNet engineers to anticipate what should be the appropriate evolution of the infrastructure to pass the next spring peak (Figure~\ref{fig:plantnet_new_users}) without problems and also to know what should be done the following years. 
    
    %\item A \textbf{large scale experimental validation on the Grid'5000~\cite{bolze2006grid} testbed} which highlights ...XXXX % list directly the findings and highlight their genericity for other apps
    % \begin{enumerate}
    %     \item What is the optimal software configuration (thread pool size of http, download, simsearch, and extraction tasks) for a given hardware configuration (based on the number of processors in the machine) that minimizes the user response time? How many users it can serve simultaneously?
    %     \item How does the number of simultaneous queries impact on the identification processing time? Where is the bottleneck in the processing steps? %Which processing step lasts most?
    %     \item What are the resource consumption characteristics of the Pl@ntNet Identification Engine?
    % \end{enumerate}
\end{enumerate}

\section{Background}
\label{sec:background}
%PV: introduce the section with one sentence

%\subsection{Challenges for Edge-to-Cloud Deployments}

Application workflows that need to be deployed 
%in a distributed manner 
across Edge-to-Cloud infrastructures usually
%PV: the subject is "workflows", it should be "developers"
have to configure their software components while considering various infrastructure constraints. 
%Such configurations impact performance.
% hardware and software components with different constraints and configuration parameters that determine their behaviour and consequently their performance.
For instance, in the Cloud, configurations can include compute and storage configurations, the number of topics in a data ingestion system, reserved memory in data processing frameworks, the inter-cloud network latency, \emph{etc.} In the Fog, they can include the streaming window size on gateways, the network latency and bandwidth between Fog devices, among others. In the Edge we can refer to device capabilities, the frequency of data emission, the power consumption, among others. 
%Furthermore, the interconnection capabilities vary between the Edge, Fog and Cloud due to the characteristics of those networks. Besides these hardware and software configuration possibilities, such applications are subject to other requirements like Quality of Service (QoS), Quality of Experience (QoE), security and privacy.
These environment settings and configuration parameters are 
extremely vast and their combination of possibilities virtually unlimited. Hence, the process of searching the ideal deployment and configuration of those real-life applications is challenging given the search space complexity: bad choices may result in increased financial expenses during deployment and production phases, decreased processing efficiency and poor user experience.

\subsection{A real-life application: Pl@ntNet}
\label{subsec:plantnet}
% general description of plantnet
% zoom in identification engine: how it works; processing steps; thread pools description and size;
% current performance status (maximum number of simultaneous requests it can serve [3-4s user response time]) according to the given HW config (40 processors) and SW config (threads pools size).
% research question about the optimal SW configuration to reduce user response time. Is it possible to reduce this time (adapting the SW config.)? and serve more clients?

% Pl@ntNet computing infrastructure is provided by Inria~\cite{inria-platnet}, Meso@LR~\cite{meso-platnet}, GENCI~\cite{genci-platnet}, and CIRAD~\cite{cirad-platnet}. The Pl@ntNet botanical data allows scientists to conduct research in different domains, such as ecology, agronomy, computer science, etc.

Pl@ntNet is a participatory application and platform dedicated to the production of botanical data and plant identification. Currently, the data consists of +35K plant species collected from more than 200 countries. As illustrated in Figure~\ref{fig:plantnet-geographic-scope}, using the Pl@ntNet mobile application, users (located in the Edge) may identify plants from pictures taken by their phones. Before sending these pictures, some preprocessing is done to reduce the image size.

% \begin{figure}[t]
%   \centering
%   \includegraphics[width=.78\linewidth]{fig_plantnet-geographic-scope.png}
%   \caption{Geographical distribution of Pl@ntNet observations~\cite{plantnet-observations}}
%   \label{fig:plantnet-geographic-scope}
% \end{figure}

% Using the Pl@ntNet mobile application (+10M downloads), users may identify plants from pictures taken by their phones. The Pl@ntNet  \textbf{Identification Engine}, subject to analysis in this work, is responsible for the automatic identification of species through Deep Learning. In a nutshell, the Identification Engine performs two main activities: \textit{(1) Species prediction:} refers to the feature extraction and classification of user's images; and \textit{(2) Similarity Search:} searches for the images of the botanical databases that are the most similar to the user's images. At the end of the processing, the Identification Engine returns to the user the ranked list of most probable species with their respective most similar plant pictures, allowing interactive validation by the users.

Then, the Pl@ntNet \textbf{Identification Engine} (located in the Cloud), subject to analysis in this work, is responsible for the automatic identification of species through Deep Learning. In a nutshell, the Identification Engine performs two main activities: \emph{(1) Species prediction:} refers to the feature extraction and classification of user images; and \emph{(2) Similarity Search:} searches for the images of the botanical databases that are the most similar to the user images. At the end of the processing, the Identification Engine returns the ranked list of most probable species with their respective, most similar plant pictures, allowing interactive validation by the users.

The processing performance of the Identification Engine strongly depends on the \textbf{thread pool size} configured to process the various \textbf{tasks} involved during the identification of users images. Table~\ref{tbl:task-description} presents the execution order of all tasks, the thread pool they belong to, and in which hardware they take place. Table~\ref{tbl:thread-pool-description} describes the role of each thread pool and an example of configuration currently used in the Pl@ntNet production servers. This configuration was defined by Pl@ntNet engineers based on their best practical experience with the Pl@ntNet system considering mainly the following: (a) for thread pools using CPU: a machine with 40 CPU cores available; and (b) for the GPU thread pool: the maximum number of threads which fit in GPU memory.

\begin{figure}[t]
  \centering
  \includegraphics[width=0.99\linewidth]{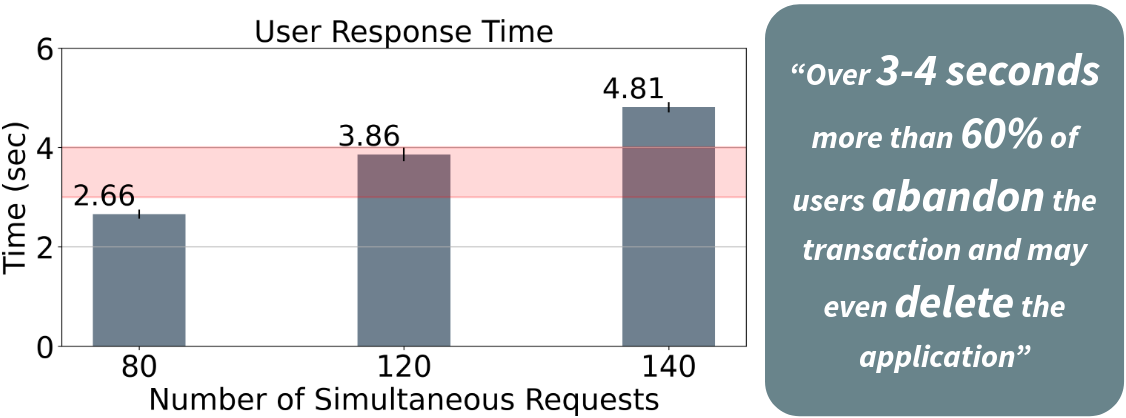}
  \caption{Pl@ntNet Engine: user response time.}
  \label{fig:response_time}
\end{figure}

The main performance metric for this application is the \textbf{user response time}. 
%Previous studies~\cite{appdynamics-report} showed that while some mobile app users may tolerate slower response times, their vast majority (60\% or more) are ready to drop the interaction and delete the mobile app altogether when experiencing response times exceeding 3-4 seconds.
A preliminary analysis~\cite{appdynamics-report} showed that to achieve a 4 seconds response time (the maximum tolerated by users) the thread pool and hardware configurations can not serve more than 120 simultaneous requests (3.86$\pm0.13$), as shown in Figure~\ref{fig:response_time}. In this context, meaningful questions that arise are: \emph{Is there a better thread pool allocation that minimizes the user response time? How many more users can the system serve if we find a better thread pool configuration?} The answers to those questions and more analytical insights will be presented in Section~\ref{sec:evaluation} through the use of our proposed methodology and its implementation in the \textbf{E2C}\textit{lab} framework.

\begin{table}[t]
\scriptsize	
% \small
\centering
\caption{Identification processing steps.}
\label{tbl:task-description}
\begin{tabular}{m{1.6cm}m{3.1cm}m{1.1cm}m{1cm}}
\hline
\textbf{Task}  & \textbf{Description}  & \textbf{Thread pool}  & \textbf{Hardware}                                              \\ \hline
\rowcolor[HTML]{E0EBEA} 
pre-process    & Decoding the query parameters. & HTTP  & CPU                                     \\
wait-download  & Wait for an available download thread. & HTTP, Download   & CPU                            \\
\rowcolor[HTML]{E0EBEA} 
download       & Download images.  & Download  & CPU                                                  \\
wait-extract   & Wait for an available extractor thread.  & HTTP, \hspace{0.5cm} Extract  & CPU, \textbf{GPU}                           \\
\rowcolor[HTML]{E0EBEA} 
extract        & DNN inference of the image.  & Extract  & \textbf{GPU}                                       \\
process        & Process classification and similarity search output at query level.  & HTTP & CPU \\
\rowcolor[HTML]{E0EBEA} 
wait-simsearch & Wait for an available similarity search thread.  & HTTP, Simsearch & CPU    \\
simsearch      & Search the most similar images in our database.  & Simsearch  & CPU                   \\
\rowcolor[HTML]{E0EBEA} 
post-process   & Check processed query results and format the response.  & HTTP & CPU             \\ \hline
\end{tabular}

\vspace{0.5cm}

\caption{Thread pool configuration of Pl@ntNet Engine.}
\label{tbl:thread-pool-description}
\begin{tabular}{m{1.4cm}m{1.2cm}m{3.3cm}m{1cm}}
\hline
\textbf{Thread pool} & 
% \textbf{Size \\(\emph{\#} threads)}  
\textbf{\begin{tabular}[c]{@{}l@{}}Size \\(\emph{\#} threads)\end{tabular}}
% \textbf{\begin{tabular}[c]{@{}l@{}}refined\\ optimum\end{tabular}}
& \textbf{Description}     & \textbf{Hardware}                          \\ \hline
\rowcolor[HTML]{E0EBEA} 
HTTP                 & 40                     & \emph{\#} simultaneous requests being processed. & CPU   \\
Download             & 40                     & \emph{\#} simultaneous images being downloaded. & CPU    \\
\rowcolor[HTML]{E0EBEA} 
Extract              & 7                      & \emph{\#} simultaneous inferences in a single GPU. & \textbf{GPU} \\
Simsearch            & 40                     & \emph{\#} simultaneous similarity search.  & CPU         \\ \hline
\end{tabular}
\end{table}

% As aforementioned, this configuration can serve at maximum 60 simultaneous requests (users) for a user response time below three seconds (call for Figure~\ref{fig:response_time}). In this context, meaningful questions that arise are: \emph{Is there a better thread pool allocation that minimizes the user response time? How many more users can the system serve if we find a better thread pool configuration?} The answers to those questions and more analytics insights will be presented in Section~\ref{sec:evaluation} through the use of our proposed methodology and its implementation in the \textbf{E2C}\textit{lab} framework.

Let us highlight that Pl@ntNet is representative of other applications in the context of the Computing Continuum. As illustrated in Figure~\ref{fig:plantnet-geographic-scope}, it consists of many geographically distributed devices (over 10 million users) that collect and send data (about 400K plant images per day), and perform preprocessing at the Edge, followed by extensive processing (\emph{e.g.}, species prediction, similarity search, \emph{etc.}) in centralized Cloud/HPC infrastructures.

\subsection{Formalizing deployment optimization on the Edge-to-Cloud Continuum}
% \begin{todo}
% [Alex] A formal definition of the optimization in this context is needed here. 
% \end{todo}
% what do we understand by optimization, some examples, metrics to optimize, (general description, take a look at the problem statement of papers)

We describe our optimization problem by defining: the \textbf{optimization variables}, the \textbf{objective function}, and the \textbf{constraints} (Equation~\ref{eq:optim_problem}). 
%The mathematical formulation of an optimization problem is presented in Equation~\ref{eq:optim_problem}, as one may note, the \emph{objective function} and the \emph{constraints} are functions of the \emph{optimization variables}~\cite{bhatti2000optimization}.

\begin{equation}
\scriptsize	
\label{eq:optim_problem}
\begin{aligned}
& \underset{x}{\text{min/max}}
& & f_m(x), \;\;\;\;\;\;\;\;\;\;\;\;\;\; m = 1, 2, \ldots, M \\
& \text{subject to}
& & g_j(x) \ \leq 0, \;\;\;\;\;\;\;\; j = 1, 2, \ldots, J \;\;\;\;\; Inequality \; constraints. \\
&&& h_k(x) \ = 0, \;\;\;\;\;\;\;\; k = 1, 2, \ldots, K \;\;\; Equality \; constraints. \\
&&& x^L_i \ \leq x_i \ \leq x^U_i, \; i = 1, 2, \ldots, I \;\;\;\; Bounds \; on \; variables.
\end{aligned}
\end{equation}

\begin{figure}[t]
  \centering
  \includegraphics[width=\linewidth]{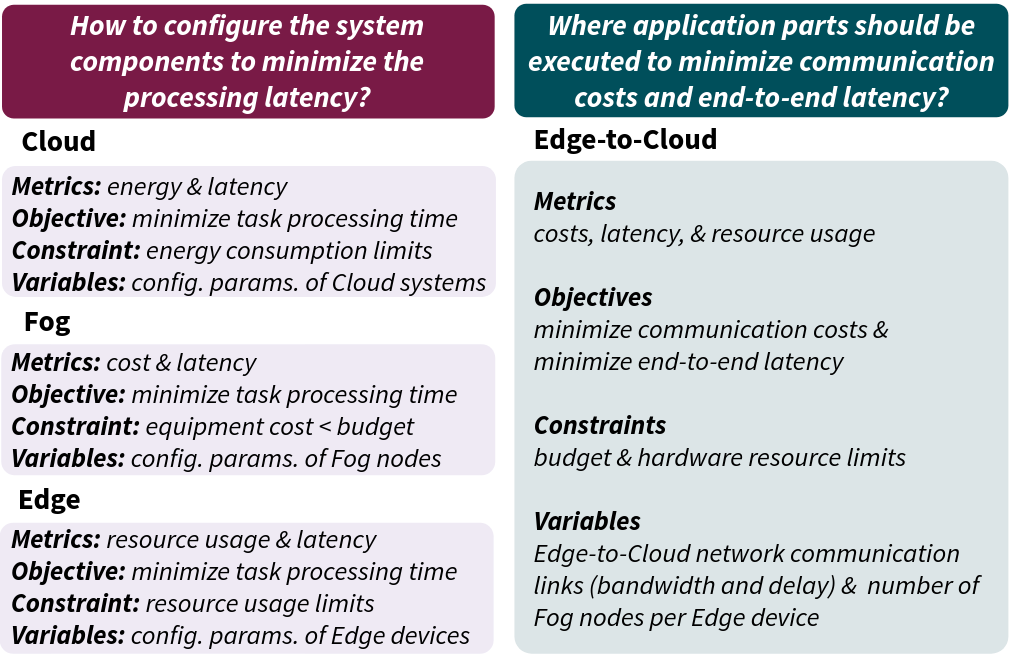}
  \caption{Edge-to-Cloud Continuum optimization problems.}
  \label{fig:optimization-edge-to-cloud}
\end{figure}

Typically, Edge-to-Cloud deployment optimization problems aim at optimizing metrics~\cite{bellendorf2020classification, aslanpour2020performance} related to: performance (\emph{e.g.}, execution time, latency, and throughput), resource usage (\emph{e.g.}, GPU, CPU, memory, storage, and network), energy consumption, financial costs, and quality attributes (\emph{e.g.}, reliability, security, and privacy). Therefore, regarding the formulation of an optimization problem and its mathematical representation in Equation~\ref{eq:optim_problem}, the \textbf{optimization variables} $x$ refer to the variables associated with the optimization problem (\emph{e.g.}, storage capacity of Edge devices, or number of cores on Fog nodes).

The \textbf{objective function} refers to the optimization objective, such as minimizing or maximizing a given metric or set of metrics (\emph{e.g.}, performance, energy consumption). The objective function maps the values of the optimization variables onto real numbers and may be classified as single-objective (such as minimizing Edge-to-Cloud processing latency) or multi-objective (\emph{e.g.}, minimizing energy consumption of Fog nodes and maximize throughput).

Finally, the \textbf{constraints} refer to requirements that a given solution must satisfy. Constraints may refer to a specific optimization variable (\emph{e.g.}, number of cores on Fog nodes between 10 and 20) and the metrics to be optimized by the objective function (\emph{e.g.}, the maximum response time must be less than 3 seconds). 

% \footnote{With an impact on accuracy discussed in Section~\ref{sec:phase1}}

Figure~\ref{fig:optimization-edge-to-cloud} depicts some examples of optimization problems. Left, one would like to answer the question: \emph{how to configure the system components to minimize processing latency?} %In this example, in order
To reduce complexity, the optimization problem is divided into three sub-problems each one with the objective of minimizing the task processing time on the Edge, Fog, and Cloud infrastructures, under specific constraints. 
%The first sub-problem (Cloud) aims at finding the configuration parameters of Cloud systems and has as constraints the energy consumption on those systems to process the task. The second sub-problem (Fog) aims at finding the configuration parameters of a cluster of Fog nodes subject to a budget limit. The third sub-problem (Edge) aims at finding the configuration parameters of Edge devices while respecting their resource constraints.
The right-hand example aims at answering the question: \emph{where should the workflow components be executed to minimize communication costs and end-to-end latency?} This translates into a single multi-objective optimization problem (minimizing communication costs and end-to-end latency), as opposed to the previous example (several single-objective optimization problems).
%The constraints refer to the available budget and the hardware resource limits. The goal is to determine the Edge, Fog, and Cloud network configurations (bandwidth and delay) and the size of the Fog cluster to process the data from Edge devices.

In order to model and solve such optimization problems, one may find multiple methods and packages in the literature. For instance, packages and libraries such as \emph{Scikit-Optimize}~\cite{scikit-optimize}, \emph{Scikit-Learn}~\cite{pedregosa2011scikit}, \emph{Surrogate Modeling Toolbox (SMT)}~\cite{SMT2019}, \emph{DeepHyper}~\cite{balaprakash2018deephyper}, etc., may be used to build surrogate models and then use those model to explore the search space of the optimization problem.

\subsection{\textbf{E2C}\textit{lab}: reproducible Edge-to-Cloud experiments}

% \textcolor{blue}{
\textbf{E2C}\textit{lab}~\cite{rosendo2020e2clab} is a framework that implements a rigorous methodology for designing experiments with real-world workloads on the Edge-to-Cloud Computing Continuum. This methodology, illustrated in Figure~\ref{fig:methodology}, provides guidelines to move from real-world use cases to the design of relevant testbed setups for experiments enabling researchers to understand performance and to support the reproducibility of the experiments.
% }

% \textcolor{blue}{
\textbf{E2C}\textit{lab} architecture is described in Figure~\ref{fig:e2clab_arch}. The idea is that experiments can accurately reproduce relevant behaviors of a given application workflow on representative settings of the physical infrastructure underlying this application. 
% }

% \textcolor{blue}{
The key features provided by \textbf{E2C}\textit{lab} are: (1) reproducible experiments; (2) the mapping of applications parts executed across the computing continuum with the physical testbed; (3) the support for experiment variation and transparent scaling of the scenario; (4) network emulation to define Edge-to-Cloud communication constraints; and (5) experiment deployment, monitoring and backup of results. \textbf{E2C}\textit{lab} is open source and is available at~\cite{e2clab-code}.
% }

\section{A Methodology for Optimizing the Performance of Applications on the Edge-to-Cloud Continuum}
\label{sec:methodology}

\begin{figure}[t]
  \centering
  \includegraphics[width=\linewidth]{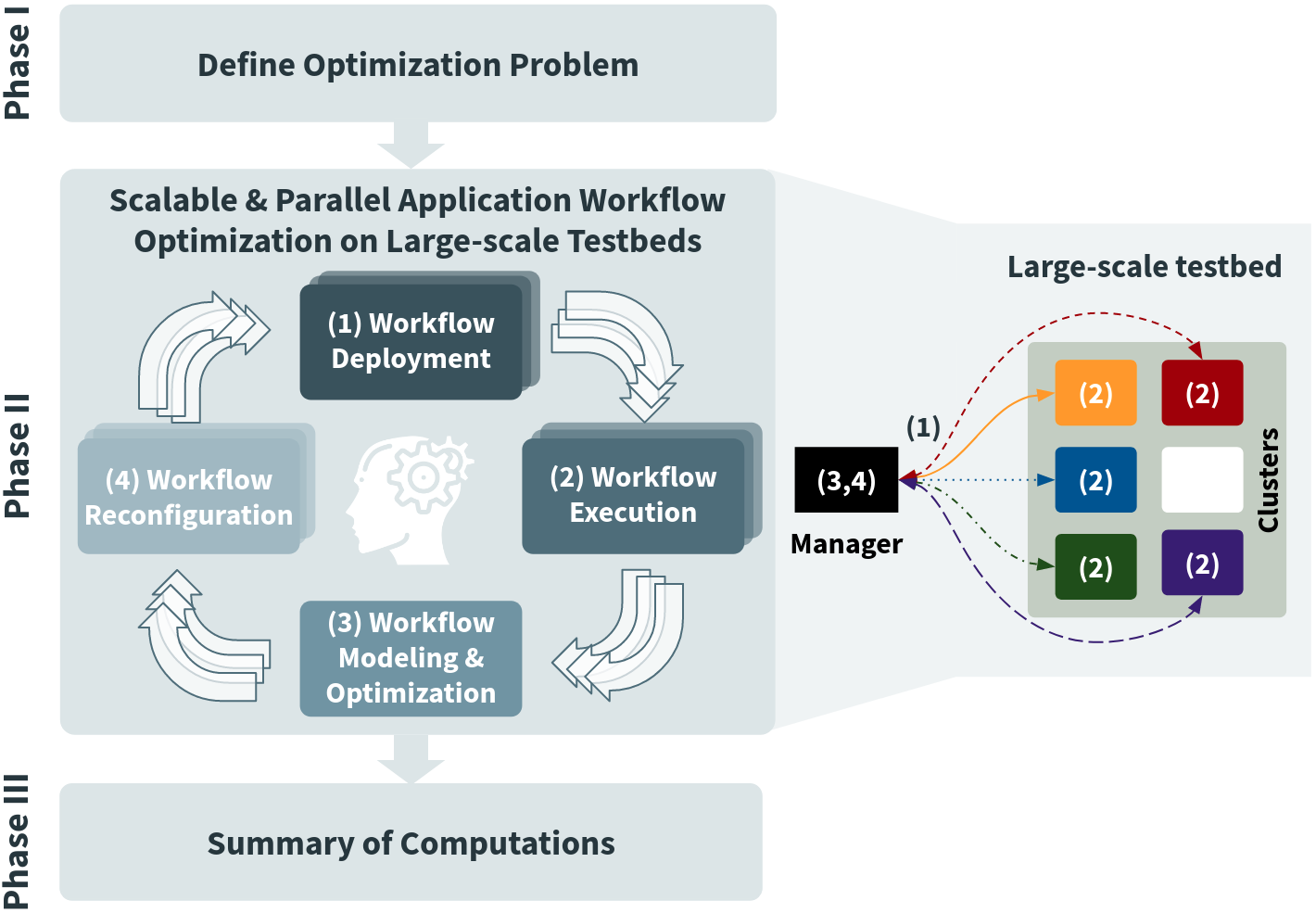}
  \caption{Our proposed optimization methodology.}
  \label{fig:optimization_details}
\end{figure}

Our optimization methodology supports reproducible parallel optimization of application workflows on large-scale testbeds. It 
%goal is to support the deployment \textbf{optimization} of application workflows executed across the Edge-to-Cloud Continuum.
%PV: the subject is "our goal"??
consists of three main phases %\textit{Initialization}, \textit{Evaluation}, and \textit{Finalization} 
illustrated in Figure~\ref{fig:optimization_details}.
%that we describe in detail in the next subsections.

\subsection{Phase I: Initialization}
\label{sec:phase1}
This phase, depicted at the top of Figure~\ref{fig:optimization_details}, consists in defining the optimization problem. The user must specify: the \textbf{optimization variables} that compose the search space to be explored (\emph{e.g.,} GPUs used for processing, Fog nodes in the scenario, network bandwidth, \emph{etc.}); the \textbf{objective} (\emph{e.g.,} minimize end-to-end latency, maximize Fog gateway throughput, \emph{etc.}); and \textbf{constraints} (\emph{e.g.} the upper and lower bounds of optimization variables, budget, response time latency, \emph{etc.}).

%Thanks to the \textit{Layers and Services} abstraction provided by our methodology, 
One may focus the optimization on: (1) specific parts of the infrastructure (\emph{e.g.,} only on geographically distributed Edge sites, or only on Fog-to-Cloud resources) by defining multiple, per infrastructure, optimization problems, as presented in the left side of Figure~\ref{fig:optimization-edge-to-cloud}. This approach reduces the search space complexity (in case of use cases with large search spaces) and hence the computing time; (2) or the whole Edge-to-Cloud infrastructure as a single optimization problem, as presented in the right side of Figure~\ref{fig:optimization-edge-to-cloud}.

% Define the Search Space
    % Thread pool size according to the number of cores on the host.
        % http_pool = (0.5*n_cores, 1.5*n_cores)
        % download_pool = (0.5*n_cores, 1.5*n_cores)
        % simsearch_pool = (0.5*n_cores, 1.5*n_cores)
        % extrac_pool = (0.5*n_cores, 1.5*n_cores)
% Define the Objective
    % minimize the identification processing time
% Define the constraints

\subsection{Phase II: Evaluation}
This phase aims at defining the mathematical methods and optimization techniques used in the \emph{optimization cycle} (presented in the middle of Figure~\ref{fig:optimization_details}) to explore the search space. Such \emph{optimization cycle} consists in: (1) parallel deployment of the application workflow in a large-scale testbed; (2) their simultaneous execution; (3) asynchronous model optimization; and (4) reconfiguration of the application workflow for a new evaluation.

This cycle continues until model convergence. Depending on the run time characteristics of the application workflows, their evaluations may be performed differently.

\subsubsection{Long-time Running Applications} refer to experiments or simulations for which the evaluation of a single point in the search space requires a lot of time to complete (\emph{e.g.,} hours, or even days). Furthermore, since application workflows in the context of the Computing Continuum typically consist of cross-infrastructure parameter configurations resulting in a myriad of configuration possibilities, their optimization problem presents a complex and large search space.

For those long-time running applications, a variety of Bayesian Optimization~\cite{snoek2012practical} methods (\emph{e.g.,} surrogate models as: Gaussian process (Kriging)~\cite{simpson2001kriging}, Decision Trees~\cite{wang2000optimization}, Random Forest~\cite{breiman2001random}, Gradient Boosting Regression Trees~\cite{friedman2001greedy}, Support Vector Machine~\cite{steinwart2008support}, Polynomial Regression~\cite{ostertagova2012modelling}, among others) may be applied as candidates to explore the search space. Their generation is described below.

\textbf{Surrogate Model Building:} this consists of three steps: \textit{(a)} a few sample points are generated, respecting the upper and lower limits of each optimization variable that composes the search space. Sampling methods such as Latin Hypercube Sample~\cite{helton2003latin} or Low Discrepancy Sample~\cite{kocis1997computational} may be applied; \textit{(b)} then, from the generated sample, parallel experiments (deployment of application workflows) are run for each parameter set; \textit{(c)} lastly, the surrogate model is trained on the dataset generated in the previous step.

\textbf{Model Retraining \& Application Optimization:} once the surrogate model is trained on the sample points previously generated, it is used to explore the optimization search space by deciding the subsequent application configurations to be evaluated in parallel. As soon as the evaluations finish, the model is retrained and optimized asynchronously, then new points are suggested to be evaluated.

\subsubsection{Short-time Running Applications} refer to the case when a few minutes are enough to evaluate a single point in the search space. Such applications also follow the \emph{optimization cycle} previously presented. Besides, they may also use surrogate models to explore the search space. However, differently from \emph{Long-time Running Use Cases}, they can use other optimization techniques such as evolutionary algorithms and swarm intelligence based algorithms (\emph{e.g.,} Genetic Algorithm~\cite{mirjalili2019genetic}, Differential Evolution~\cite{das2016recent}, Simulated Annealing~\cite{van1987simulated}, Particle Swarm Optimization~\cite{du2016particle}, \emph{etc.}).

\subsection{Phase III: Finalization}
For \textbf{reproducibility} purposes, this last phase illustrated at the bottom of Figure~\ref{fig:optimization_details} provides a summary of computations. Therefore, it provides: the definition of the optimization problem (optimization variables, objective, and constraints); the sample selection method; the surrogate models or search algorithms with their hyperparameters used to explore the search space of the optimization problem; and finally the best application configuration found. Providing all this information at the end of computations allows other researches to reproduce the research results.

\begin{figure}[t]
    \centering
         \includegraphics[width=0.96\linewidth]{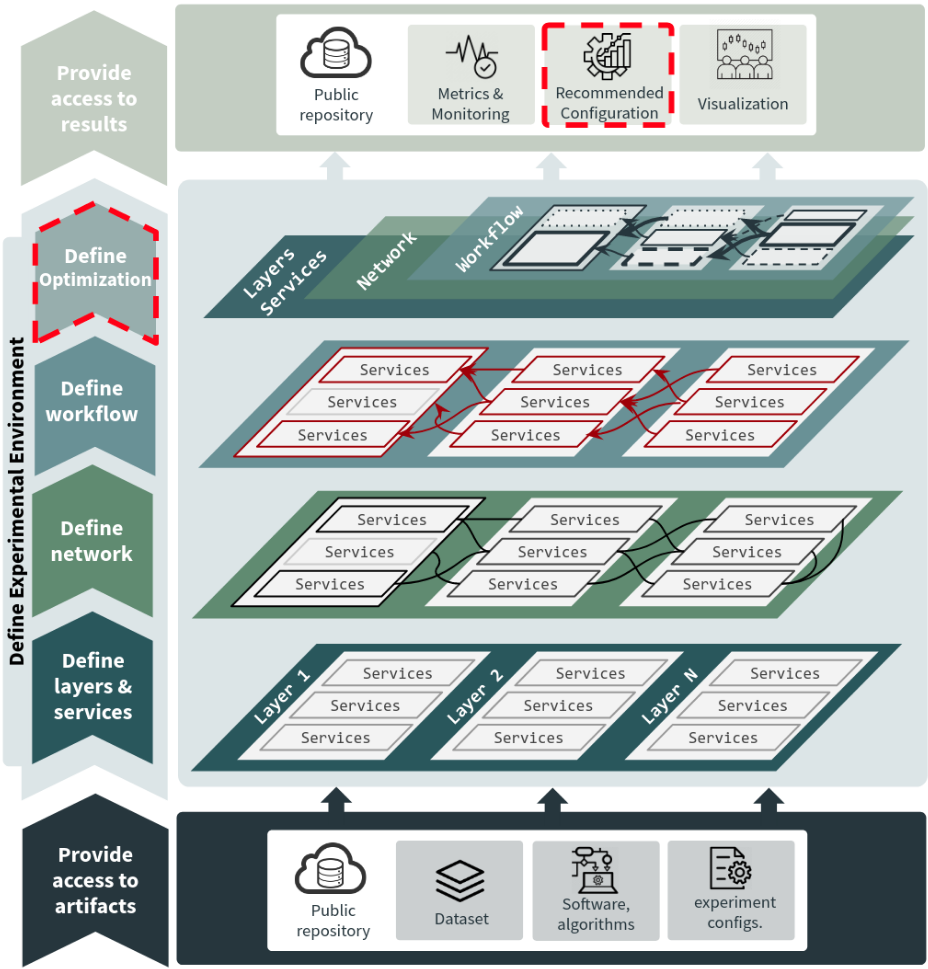}
    \caption{Extended \textbf{E2C}\textit{lab} experimental methodology.}
    \label{fig:methodology}
\end{figure}

\subsection{Implementation as an extension of the \textbf{E2C}\textit{lab} framework}
\label{sec:methodology_impl}

To validate our optimization approach, we enhanced the \textbf{E2C}\textit{lab} framework for reproducible experimentation across the Edge-to-Cloud Continuum. We extended~\cite{e2clab-code} the \textbf{E2C}\textit{lab} framework~\cite{e2clab-doc-page} with support for the performance optimization of application workflows. Figure~\ref{fig:methodology} shows a holistic view of our enhanced methodology containing the extensions highlighted in dashed lines colored in red. As one may note, we have added a new sub-process named \textit{Define Optimization} (detailed in Figure~\ref{fig:optimization_details}) inside the \textit{Define the Experimental Environment} process. % The \textit{Define Optimization} sub-process contemplates the three previous sub-processes meaning that one may optimize all the elements within the Edge-to-Cloud Continuum, such as \textit{Layers and Services}, \textit{Network}, and application \textit{Workflow}. 

\begin{figure}[t]
    \centering
         \includegraphics[width=0.81\linewidth]{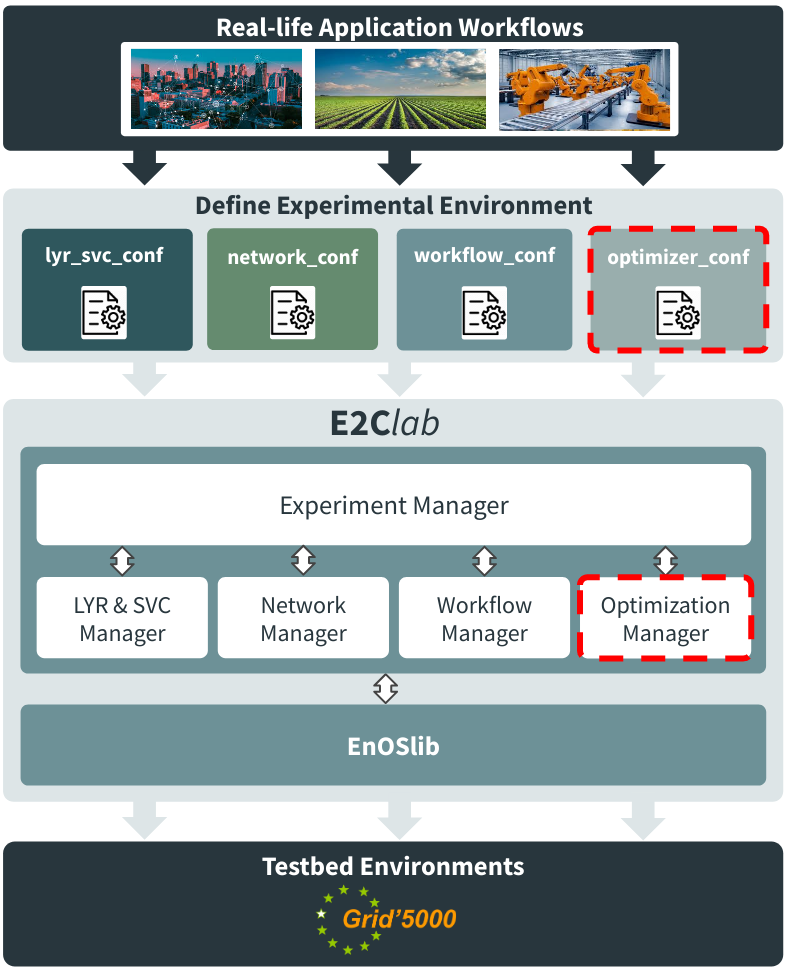}
    \caption{Extended \textbf{E2C}\textit{lab} architecture.}
    \label{fig:e2clab_arch}
\end{figure}

Figure~\ref{fig:e2clab_arch} illustrates the \textbf{E2C}\textit{lab} architecture. We designed a new manager named \emph{Optimization Manager} (which implements the optimization approach in Figure~\ref{fig:optimization_details}). Its role is to: interpret the user-defined optimization setup defined in the \emph{optimizer\_conf} configuration file and then automate the \emph{optimization cycle} (1. parallel deployment of the application workflow in a large-scale testbed; 2. simultaneous application workflow execution; 3. asynchronous model optimization; and 4. reconfiguration of the application workflow for a new evaluation) in order to optimize the application workflow. Laslty, the \emph{Optimization Manager} provides a summary of computations for reproducibility purposes. 

% \textcolor{blue}{
The \emph{Optimization Manager} takes advantage of \emph{Ray}~\cite{ray-api} to run parallel application workflows on the Grid'5000 large-scale testbed. Ray Tune~\cite{liaw2018tune} provides state of the art search algorithms; manages model checkpoints and logging; and methods for analyzing training.
% }

% \textcolor{blue}{
\textbf{User-defined optimization} (\emph{i.e., how to setup an optimization?}): the \emph{Optimization Manager} offers a class-based API that allows researchers to setup and control the model training. Users have to inherit the \emph{Optimization class} and define in the \emph{run()} function (Listing~\ref{lis-udo} line~\ref{line:run}) the optimization configuration through several state of the art single-objective and multi-objective Bayesian Optimization search algorithms (\emph{e.g.}, from libraries such as \emph{Scikit-Optimize}~\cite{scikit-optimize}, \emph{Dragonfly}~\cite{kandasamy2019tuning}, \emph{Ax}~\cite{balandat2020botorch}, \emph{HEBO}~\cite{cowen2020hebo}, among others). Next, users define in the \emph{run\_objective()} function (Listing~\ref{lis-udo} line~\ref{line:obj}) their optimization logic, which runs in parallel to train the model. To do so, the \emph{Optimization class} provides the following three methods: %(1) \textit{prepare()}, Listing~\ref{lis-udo} line~\ref{line:prepare}; (2) \textit{launch()}, Listing~\ref{lis-udo} line~\ref{line:launch}; and (3) \textit{finalize()}, Listing~\ref{lis-udo} line~\ref{line:finalize}.
% }

\begin{enumerate}
    \item \textit{prepare()}: for reproducibility of optimization evaluations, it generates a dedicated optimization directory for each model evaluation (Listing~\ref{lis-udo} line~\ref{line:prepare}).
    
    \item \textit{launch()}: deploys the application on a large-scale testbed to perform a model evaluation (Listing~\ref{lis-udo} line~\ref{line:launch}). For reproducibility, deployment-related information are captured, such as physical machines, network constraints, and application configurations.
    
    \item \textit{finalize()}: for reproducibility purposes, it stores the optimization computations for a given model evaluation in the optimization directory created in the \emph{prepare()} phase (Listing~\ref{lis-udo} line~\ref{line:finalize}). Saved information refers to intermediate models throughout training and points evaluated.
\end{enumerate}

Listing~\ref{lis-udo} shows how one may define an optimization problem (e.g., Pl@ntNet problem in Eq.~\ref{eq:optim_problem_plantnet}). A detailed example may be found on the \textbf{E2C}\textit{lab} documentation Web page~\cite{e2clab-doc-page}.

%Listing~\ref{lis-std-optim-conf} shows how one may define an optimization problem and Table~\ref{tbl:config-files} describes the attributes of the \emph{optimizer\_conf} file.

% Besides, it uses \emph{Scikit-Optimize}~\cite{scikit-optimize} for surrogate-based Bayesian optimization. \emph{Scikit-Optimize} is a library (built on top of \emph{Scikit-Learn}~\cite{pedregosa2011scikit}) for sequential model-based optimization to minimize expensive black-box functions.

% \emph{DeepHyper} is a scalable automated machine learning package for Neural Architecture Search (search for high-performing deep neural network architectures) and Hyperparameter Search for Deep Learning (hyperparameter optimization for a given model).

% describe the the whole processing; and how their work together

% \begin{todo}
% [Alex] Some example of the "implementation" is needed. For instance, explain how using the Layers and Services abstraction of E2Clab, the Define optimization layer can be instanciated (i.e., how one can define the problem in Phase 1 for instance)
% \end{todo}

% \begin{figure}[t]
%     \lstset{aboveskip=0pt,belowskip=0pt}
%     \lstinputlisting[language=Python, caption=Optimizer configuration file example., label=lis-std-optim-conf]{listing_optimization_conf}
% \end{figure}

\begin{figure}[t]
    \lstset{aboveskip=0pt,belowskip=0pt}
\begin{lstlisting}[language=Python, escapechar=|, caption=Example of a user-defined optimization in \textbf{E2C}\textit{lab}., label=lis-udo]
from e2clab.optimizer import Optimization

class UserDefinedOptimization(Optimization):

    def run(self):|\label{line:run}|
        algo = SkOptSearch(|\label{line:sab}|
            optimizer=Optimizer(
                base_estimator='ET', 
                n_initial_points=45, 
                initial_point_generator="lhs", 
                acq_func="gp_hedge"))|\label{line:sae}|
        algo = ConcurrencyLimiter(algo, max_concurrent=2)
        scheduler = AsyncHyperBandScheduler()
        objective = tune.run(
            self.run_objective,
            metric="user_resp_time",
            mode="min",
            name="plantnet_engine",
            search_alg=algo,
            scheduler=scheduler,
            num_samples=10,
            config={
                "http": tune.randint(20, 60),
                "download": tune.randint(20, 60),
                "simsearch": tune.randint(20, 60),
                "extrac": tune.randint(3, 9)})

    def run_objective(self, _config):|\label{line:obj}|
        # create an optimization directory
        self.prepare()|\label{line:prepare}|
        # deploy the configs on the testbed
        self.launch()|\label{line:launch}|
        # backup the optimization computations
        self.finalize()|\label{line:finalize}|
        # report the metric value to Ray Tune
        tune.report(user_resp_time=user_resp_time)
\end{lstlisting}
\end{figure}

\section{Experimental Validation}
\label{sec:evaluation}
In this section we illustrate our proposed optimization methodology by showing how %\textbf{E2C}\textit{lab} 
it can be used to analyze the performance of the Pl@ntNet botanical application and to find its thread pool configurations. The goal of our experiments is to answer the following research questions:

\begin{enumerate}
        \item What is the software configuration, for a given hardware configuration, that minimizes the user response time?
        \item How does the number of simultaneous users accessing the system impact on the user response time?
        \item How do the \emph{Extraction} and \emph{Similarity Search} thread pool configurations impact the processing time and user response time? % Where is the processing bottleneck of the Pl@ntNet Engine?
        % \item What are the resource consumption characteristics of the Pl@ntNet Engine?
\end{enumerate}
% \begin{enumerate}
%         \item What is the best software configuration of the Pl@ntNet Identification Engine, for a given hardware configuration, that minimizes the user response time?
%         \item How does the number of simultaneous users accessing the Pl@ntNet system impact on the processing time and user response time?
%         \item How does the \emph{Extraction} and \emph{Similarity search} thread pool configurations impact on the processing time and user response time? % Where is the processing bottleneck of the Pl@ntNet Engine?
%         % \item What are the resource consumption characteristics of the Pl@ntNet Engine?
% \end{enumerate}

% \begin{table}[t]
% \small
% \centering
% \caption{Extra trees regressor model parameters~\cite{scikit-optimize}.}
% \label{tbl:model-params}
% \begin{tabular}{ll}
% \hline
% \textbf{Parameters}       & \textbf{Values} \\ \hline
% \rowcolor[HTML]{E0EBEA} 
% n\_initial\_points        & 45             \\
% initial\_point\_generator & LHS            \\
% \rowcolor[HTML]{E0EBEA} 
% acq\_func                 & gp\_hedge      \\
% n\_points                 & 10000          \\
% \rowcolor[HTML]{E0EBEA} 
% xi                        & 0.01           \\
% kappa                     & 1.96           \\
% \rowcolor[HTML]{E0EBEA} 
% n\_calls                  & 10            
% \end{tabular}
% \end{table}

The experimental setup is defined as follows: 

\emph{\textbf{a) Scenario Configuration:}} the experiments are carried out on 42 nodes of the Grid'5000~\cite{bolze2006grid} testbed (clusters \textit{chifflot}, \textit{chiclet}, \textit{chetemi}, \textit{chifflet}, and \textit{gros}). Since the \emph{Pl@ntNet Identification Engine} requires GPU, it is deployed on the \textit{chifflot} machines (model Dell PowerEdge R740), which are equipped with Nvidia Tesla V100-PCIE-32GB GPUs, Intel Xeon Gold 6126 (Skylake, 2.60GHz, 2 CPUs/node, 12 cores/CPU), 192GB of memory, 480GB SSD, and 25Gbps Ethernet interface. The clients submitting requests to the \emph{Pl@ntNet Identification Engine} are deployed on the \textit{chiclet}, \textit{chetemi}, \textit{chifflet}, and \textit{gros} clusters. The network connection is configured with 10Gb.

\emph{\textbf{b) Workloads:}} we defined three categories of workloads, according to the number of simultaneous requests (\emph{i.e.,} 80, 120, and 140) submitted to the \emph{Pl@ntNet Identification Engine} during the whole experiment execution. 

\emph{\textbf{c) Configuration Parameters:}} Table~\ref{tbl:thread-pool-description} presents the parameters used to configure the thread pool size of the \emph{Pl@ntNet Engine}. As presented in Equation~\ref{eq:optim_problem_plantnet}, these parameters refer to the optimization variables of the optimization problem.

\emph{\textbf{d) Performance Metrics:}} the metric of interest is the \emph{user response time}. In Equation~\ref{eq:optim_problem_plantnet}, this metric is to be minimized as the optimization objective. The \emph{user response time} refers to the average time that a user waits for the response to a request. Besides this metric, we also analyze the \emph{identification processing time}, which refers to the average time to process a user request. The identification processing is divided into multiple tasks running in parallel, as described in Table~\ref{tbl:task-description}. 

We compare and analyze the \emph{user response time} and \emph{identification processing time} with respect to two thread pool configurations: \emph{\textbf{baseline}} and \emph{\textbf{preliminary optimum}}. The \emph{\textbf{baseline}} refers to the current \emph{Pl@ntNet} configuration used in the production servers. This configuration was defined by \emph{Pl@ntNet} engineers based on their best practical experience with the \emph{Pl@ntNet} system, as explained in Subsection~\ref{subsec:plantnet} and presented in Table~\ref{tbl:thread-pool-description}.

The \emph{\textbf{preliminary optimum}} configuration is the one found using our methodology. We named it preliminary since the optimization problem may have multiple \emph{minima} and one may find other application configurations if a different technique is used (\emph{e.g.,} Gaussian Process (Kriging)~\cite{simpson2001kriging}, Gradient Boosting Regression Trees~\cite{friedman2001greedy}, among others). Besides, changes in the hardware configuration (\emph{e.g.,} size of GPU memory, number of CPU cores, among others) running the Pl@ntNet application will require a new search for the thread pool sizes since their configuration strongly depends on the hardware. In this case, our optimization methodology should be applied again. In a subsequent step, we further refine the preliminary optimum using sensitivity analysis, to obtain what we call \emph{\textbf{refined optimum}} (see Section~\ref{sec:sensitivity}).

In order to obtain accurate measurements we run each experiment (each thread pool configuration) 7 times and each experiment has a duration of 23 minutes (1380 seconds). Besides, during the execution of each experiment we collect the metric values every 10 seconds. Therefore, the \emph{user response time} is presented with the mean and standard deviation regarding 966 measurements ($138*7$).

We highlight that, since through experiments we identified variations between measurements, we decided to repeat each configuration 6 times (7 experiments) to reduce the standard deviation of measurements. Besides, we run each experiment for 23 minutes with an interval of metric collection of 10 seconds to also minimize the standard deviation of the metrics collected. Furthermore, thanks to the \textbf{repeatability} feature provided in \textbf{E2C}\textit{lab}, one may repeat those experiments easily by issuing the following command: \emph{e2clab optimize --repeat 6 --duration 1380 path/to/backup/experiments/ path/to/artifacts/}.

\begin{figure}[t]
  \centering
  \includegraphics[width=0.9\linewidth]{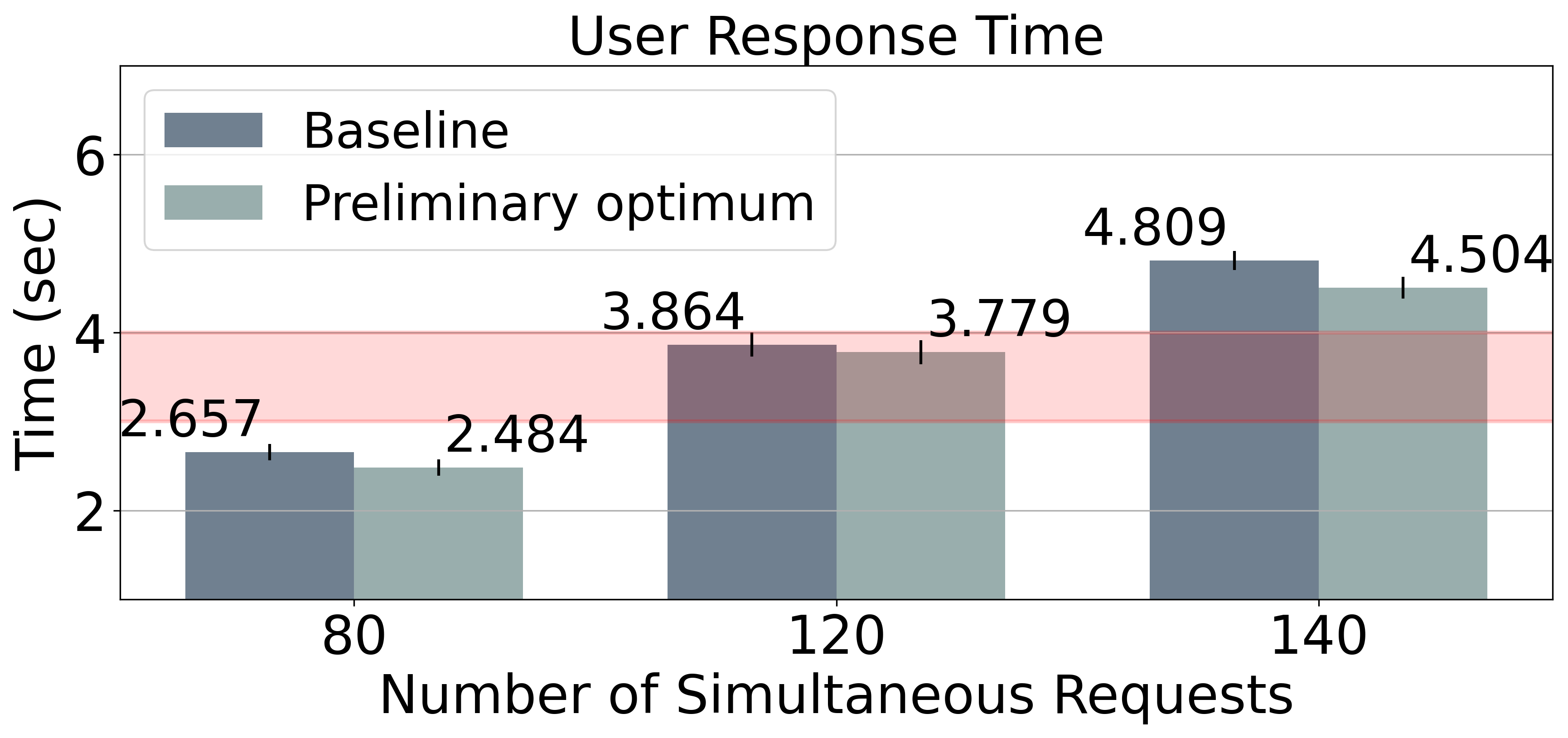}
  \caption{User response time: baseline vs preliminary.}
  \label{fig:result_baseline-vs-best-user-response-time}
\end{figure}

% \subsection{Minimizing Pl@ntNet User Response Time}
% \subsection{What is the best software configuration of the Pl@ntNet Identification Engine that minimizes the user response time?}
\subsection{What is the software configuration that minimizes the user response time?}
% describe how the models were built and preliminary optimum
% convergence plot with (response time)/configurations of (selected model) with (RF, ET, GBRT): here we show the performance of many configurations.
% table with accuracy and RMSE of (KRG, KPLS, KPLSK, MGP), highlight the best model.
% plot baseline vs best configuration (40, 60, 70):
%       user response time
%       processing times
%       resource usage: GPU (RAM, power, utilization)
%       resource usage: Energy Consumption
%       resource usage: Memory, CPU, etc.
% Identifying bottlenecks and post-optimization (maybe a preliminary optimum section?)
The optimization problem to be solved can be stated as follows:

\begin{equation}
\small
\label{eq:optim_problem_plantnet}
\begin{aligned}
% & & x = (HTTP, Download, Simsearch, Extract), \;\;\;\;\;\;\; optimization variables \\
& \text{Find} 
& & (http, download, simsearch, extract), \; in\;order\;to\\
& {\text{Minimize}} %\underset{x}{\text{min}}
& & UserResponseTime \\
& \text{Subject to}
% & & UserResponseTime  <  3\;seconds \\
& & 20 \leq (http, download, simsearch) \leq 60, \; Pool\; Size.\\
&&& \;\;3 \leq (extract) \leq 9, \;\;\;\;\;\;\;\;\;\;\;\;\;\;\;\;\;\;\;\;\;\;\;\;\;\;\;\;\;\;\;\;\; Pool\; Size.
\end{aligned}
\end{equation}

The function \emph{UserResponseTime} is given by the parallel execution of the Pl@ntNet workflow on the Grid'5000 testbed, as described in \emph{Phase II} of our methodology.

% The function \emph{UserResponseTime} is given by the surrogate model of the Pl@ntNet workflow, selected in \emph{Phase II}. Therefore, the search algorithms can solve this optimization problem evaluating a fast and accurate approximation of the Pl@ntNet workflow. Instead of deploying multiple times the entire workflow to find the best configuration, which is time consuming and computationally expensive.

In order to define the search space dimensions we run experiments to identify the maximum upper bounds of variables that do not increase the \emph{user response time} compared to the baseline Pl@ntNet configuration. Therefore, the lower and upper bounds of variables (see Equation~\ref{eq:optim_problem_plantnet}) are $\pm 50\%$ of the baseline configuration (recall Table~\ref{tbl:thread-pool-description}), respectively. 

The workload uses 80 simultaneous requests to the Pl@ntNet Identification engine. We highlight that this number has to be bigger than the upper bound of the \emph{HTTP thread pool size} since the \emph{HTTP pool} refers to the simultaneous requests being processed. 

% \textcolor{blue}{
% Scikit-optimize provides an \emph{ask} (generates the next points to evaluate) and \emph{tell} (fits the model) API that allows obtaining multiple points for evaluation in parallel. The method used to define the next points is based on a closed-form formula that allows fast and accurate deterministic approximation of \emph{q-EI} (Multi-points Expected Improvement criterion). For mathematical details on this strategy to sample multiple points, refer to~\cite{chevalier2013fast}. After the next points are defined (each one representing a thread pool configuration), they are evaluated as an experiment deployed on the testbed to obtain the \emph{UserResponseTime}. The obtained results are used to asynchronously fit the model. Next, the new model is used to suggest the next points to evaluate.
% }

We leverage Bayesian Optimization since it is typically used for global optimization of black-box functions that are expensive to evaluate~\cite{frazier2018tutorial}. \emph{Extra Trees} regressor is used as surrogate model~\cite{scikit-optimize} to model our expensive function. This surrogate model is improved by evaluating the \emph{UserResponseTime} function at the next points. The goal is to find the minimum of \emph{UserResponseTime} function with as few evaluations as possible. Listing~\ref{lis-udo} lines~\ref{line:sab} to~\ref{line:sae} detail the search algorithm parameters. The minimization has converged after 9 evaluations and the results are presented in Table~\ref{tbl:best-found} (considering a workload of 80 simultaneous requests). As one may note, the preliminary optimum configuration reduces the user response time by 7\% and can serve 35\% more simultaneous users (54 against 40, see the \emph{HTTP} thread pool).

% From the results, we highlight that thanks to our methodology implemented in \textbf{E2C}\textit{lab}, one may easily find an optimized application configuration. \textbf{E2C}\textit{lab} abstracts all the complexities to: define the whole experimental environment and the optimization problem (recall to Listing~\ref{lis-std-optim-conf}); deploy the application; run extensive parallel experiments in a large-scale testbed; and collect all the experiments results. In the next subsections we enhance our analysis in order to: (a) understand the performance of both configurations for different workloads; and (b) better understand the performance results and their correlation with the resource usage. 

From the results, we highlight that thanks to our optimization methodology implemented in \textbf{E2C}\textit{lab}, one may easily find an optimized application configuration. \textbf{E2C}\textit{lab} abstracts all the complexities to: define the whole optimization problem (recall to Listing~\ref{lis-udo}); deploy the application; run parallel evaluations of the optimization in a large-scale testbed; and collect all the experiments results. In the next subsections we enhance our analysis in order to: (a) understand the performance of both configurations for different workloads; and (b) better understand the performance results and their correlation with the resource usage.

\begin{table}[t]
\small
\centering
\caption{Baseline vs preliminary optimum configurations.}
\label{tbl:best-found}
\begin{tabular}{lll}
\hline
\textbf{Thread pool}        & \textbf{baseline}        & \textbf{\begin{tabular}[c]{@{}l@{}}preliminary \\optimum\end{tabular}}                            \\ \hline
\rowcolor[HTML]{E0EBEA} 
HTTP                        & 40                       & \cellcolor[HTML]{E0EBEA}54                      \\
Download                    & 40                       & 54                                              \\
\rowcolor[HTML]{E0EBEA} 
Extract                     & 7                        & \cellcolor[HTML]{E0EBEA}7                       \\
Simsearch                   & 40                       & 53                                              \\ \hline
% \rowcolor[HTML]{E0EBEA} 
\textbf{User response time} & \textbf{2.657 ($\pm 0.0914$)} & 
% \cellcolor[HTML]{E0EBEA}
\textbf{2.484 ($\pm 0.0912$)} \\ \hline
\end{tabular}
\end{table}

\subsection{How does the number of simultaneous users accessing the system impact on the user response time?}

In order to understand the impact of different workloads on the \emph{user response time}, we defined three workloads that represent simultaneous requests submitted to the Pl@ntNet system. The goal of these experiments is to compare the performance gains of the preliminary optimum thread pool configuration (found using our methodology) against the baseline (current Pl@ntNet configuration). Lastly, we exploit the maximum number of simultaneous requests that each configuration can handle considering the 3-4 seconds \emph{user response time} constraint.

\begin{figure*}[t]
\centering
\begin{subfigure}{.36\textwidth}
  \centering
  \includegraphics[width=\linewidth]{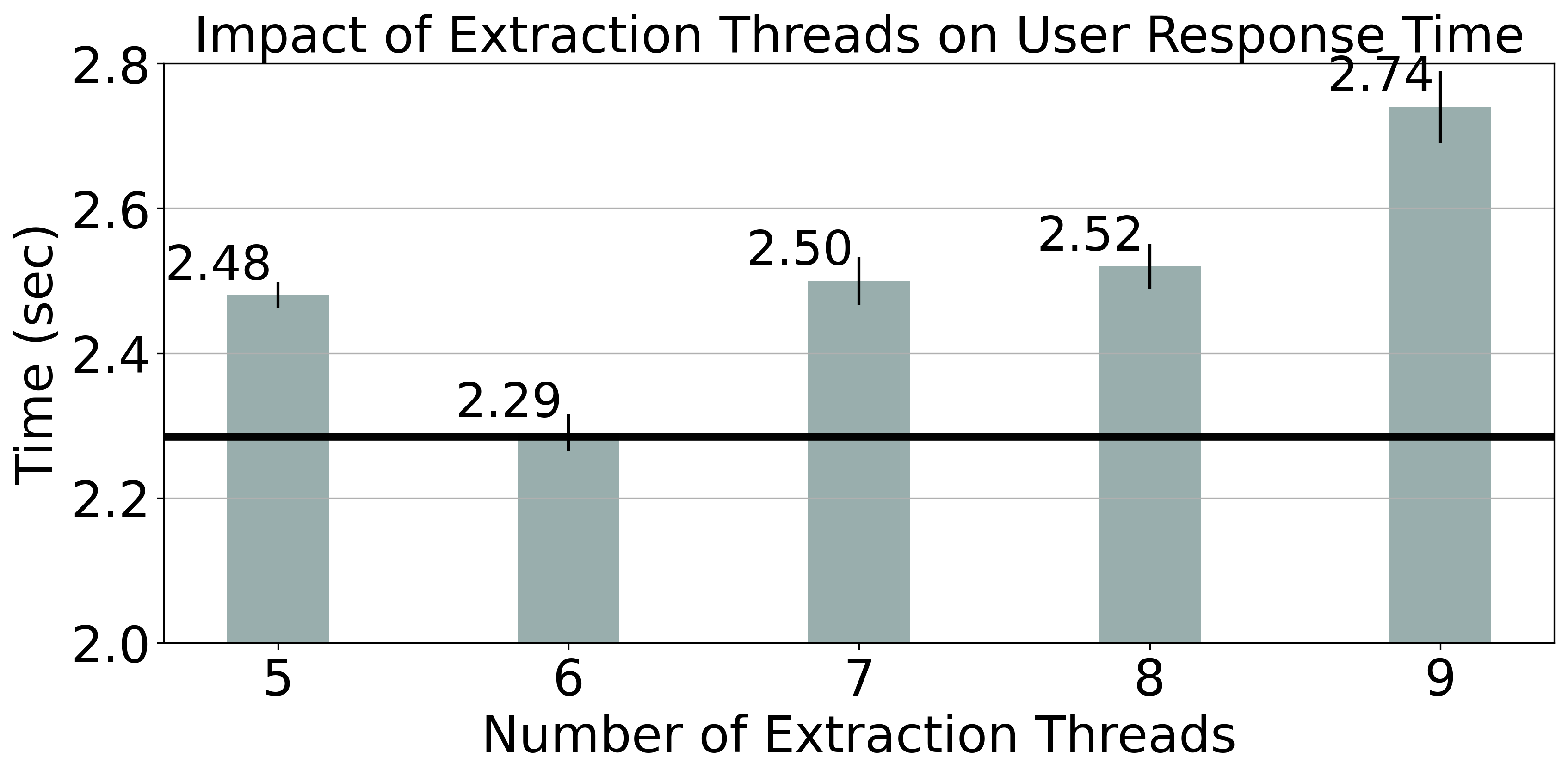}
  \caption{user response time.}
  \label{fig:result_sa_extract}
\end{subfigure}%
\begin{subfigure}{.62\textwidth}
  \centering
  \includegraphics[width=\linewidth]{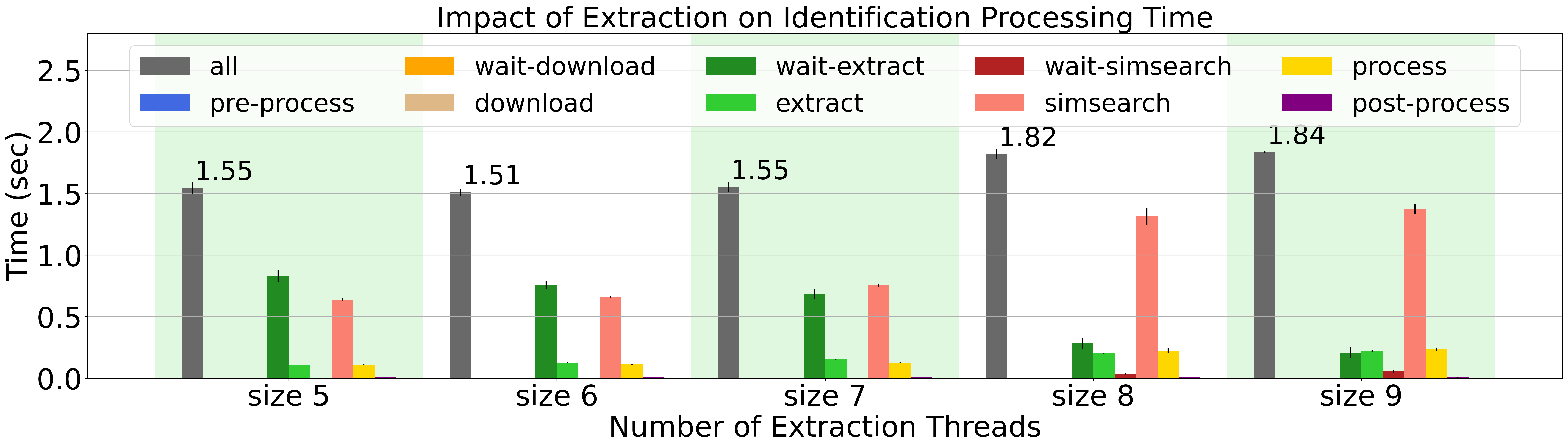}
  \caption{processing time.}
  \label{fig:result_sa_extract_pt}
\end{subfigure}

\begin{subfigure}{.309\textwidth}
  \includegraphics[width=\linewidth]{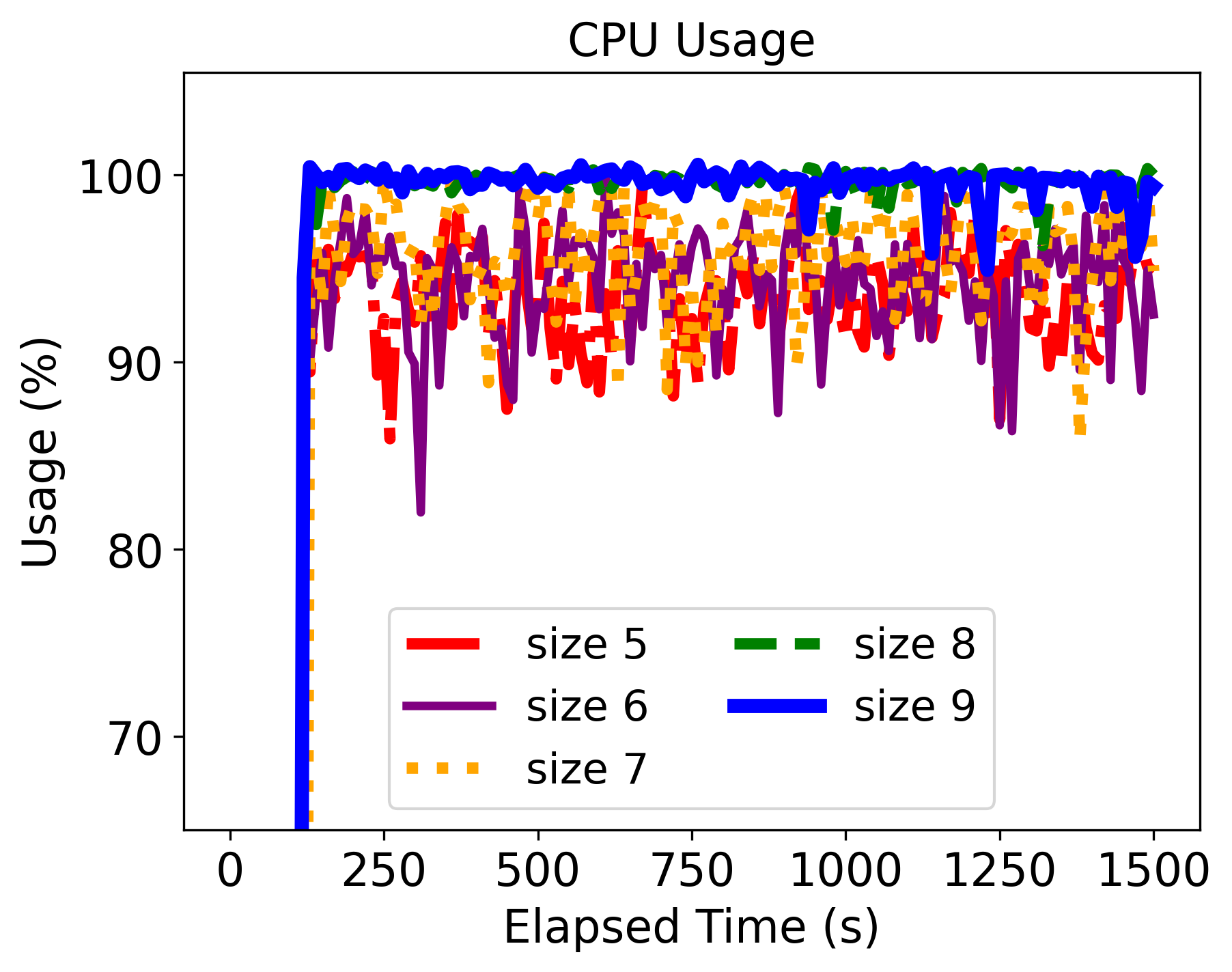}
  \caption{CPU usage.}
  \label{fig:sa_extract_cpu-usage}
\end{subfigure} \hspace{0.35cm} 
\begin{subfigure}{.309\textwidth}
  \includegraphics[width=\linewidth]{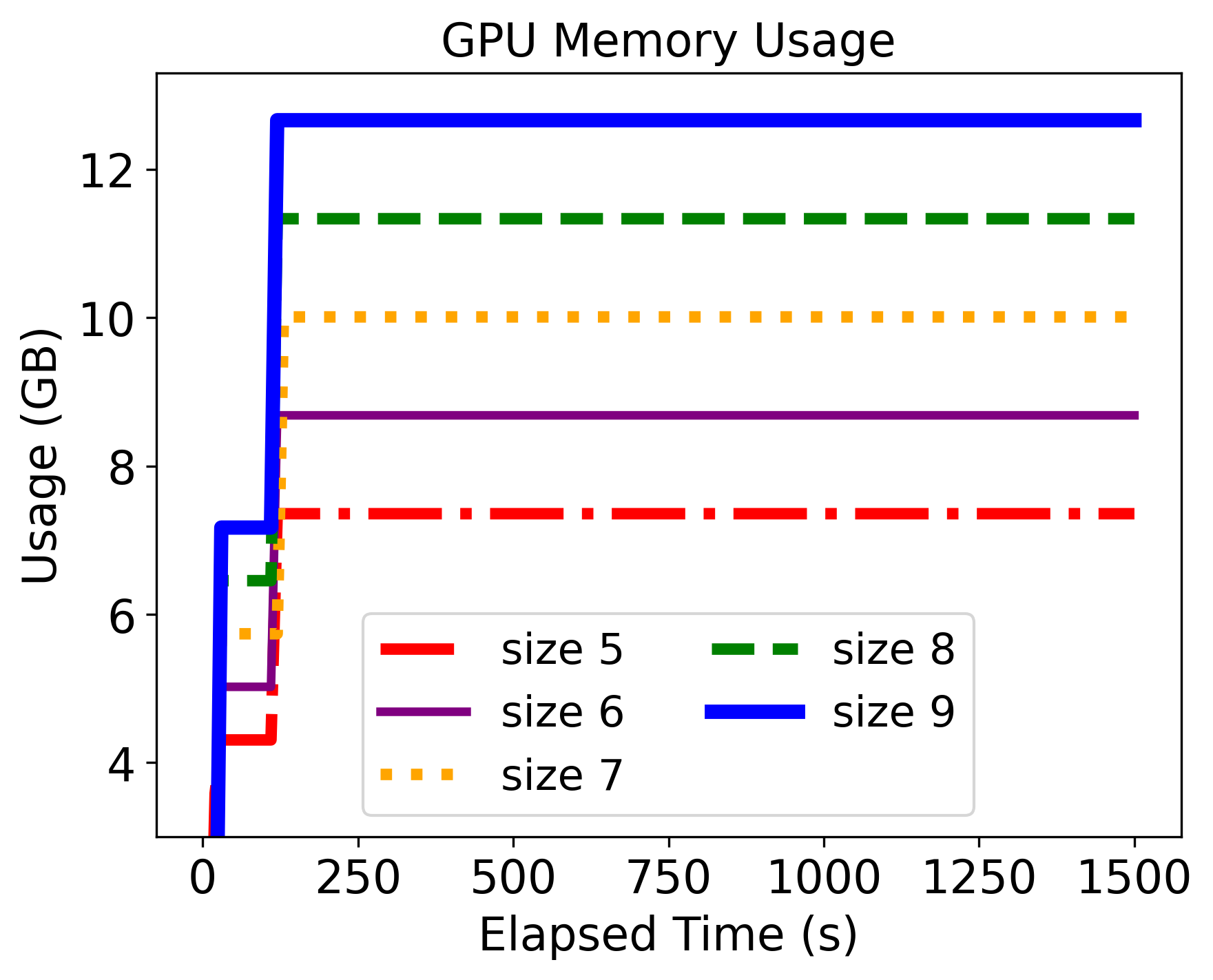}
  \caption{GPU memory usage.}
  \label{fig:sa_extract_gpu-memory-usage}
\end{subfigure} \hspace{0.35cm}
% \begin{subfigure}{.25\textwidth}
%   \centering
%   \includegraphics[width=\linewidth]{fig_sa_extract_gpu-utilization.png}
%   \caption{}
%   \label{fig:sa_extract_gpu-utilization}
% \end{subfigure}%
% \begin{subfigure}{.25\textwidth}
%   \centering
%   \includegraphics[width=\linewidth]{fig_sa_extract_gpu-power-draw.png}
%   \caption{}
%   \label{fig:sa_extract_gpu-power}
% \end{subfigure}%
\begin{subfigure}{.309\textwidth}
  \includegraphics[width=\linewidth]{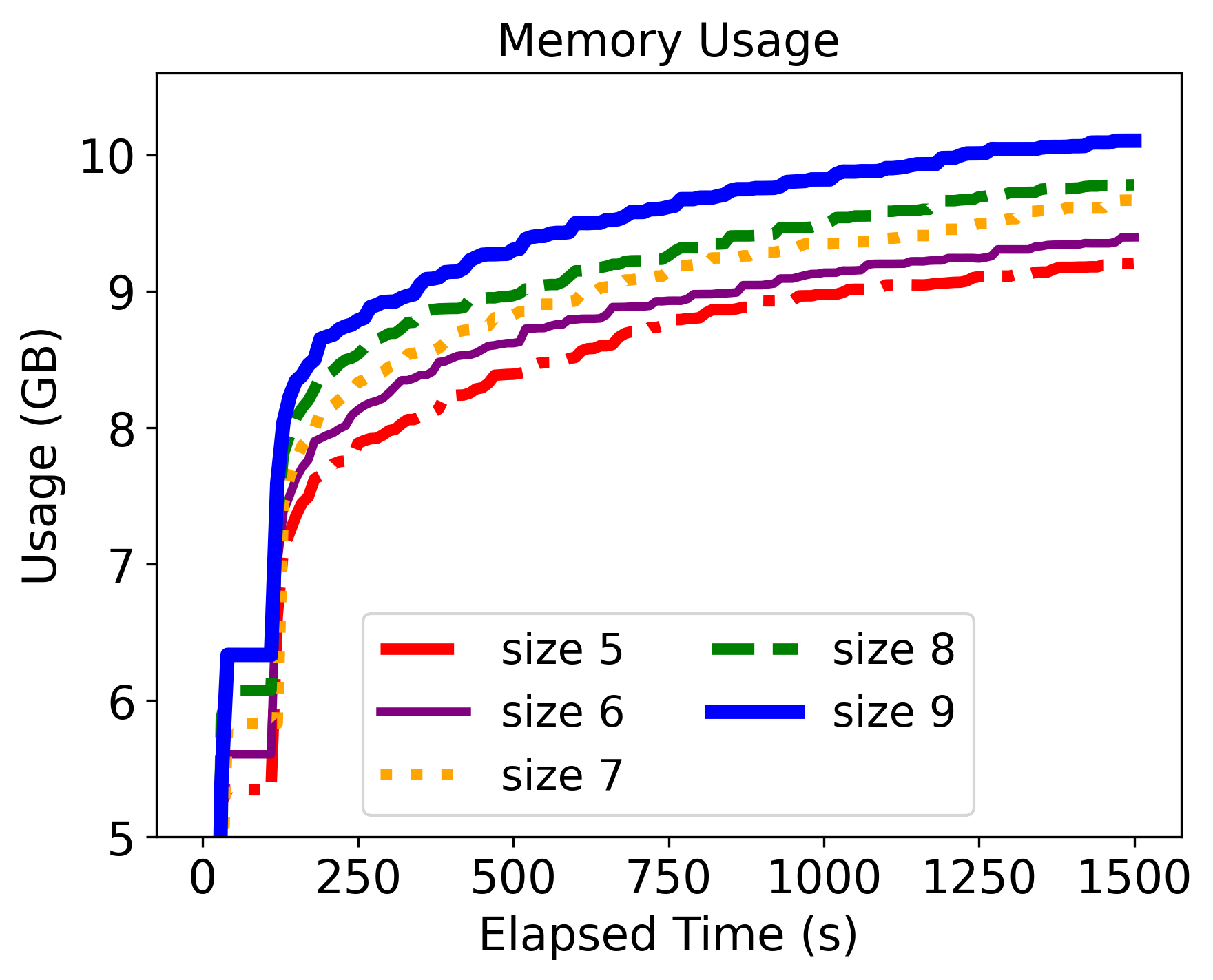}
  \caption{system memory usage.}
  \label{fig:sa_extract_memory-usage}
\end{subfigure}%

\begin{subfigure}{.49\textwidth}
  \centering
  \includegraphics[width=\linewidth]{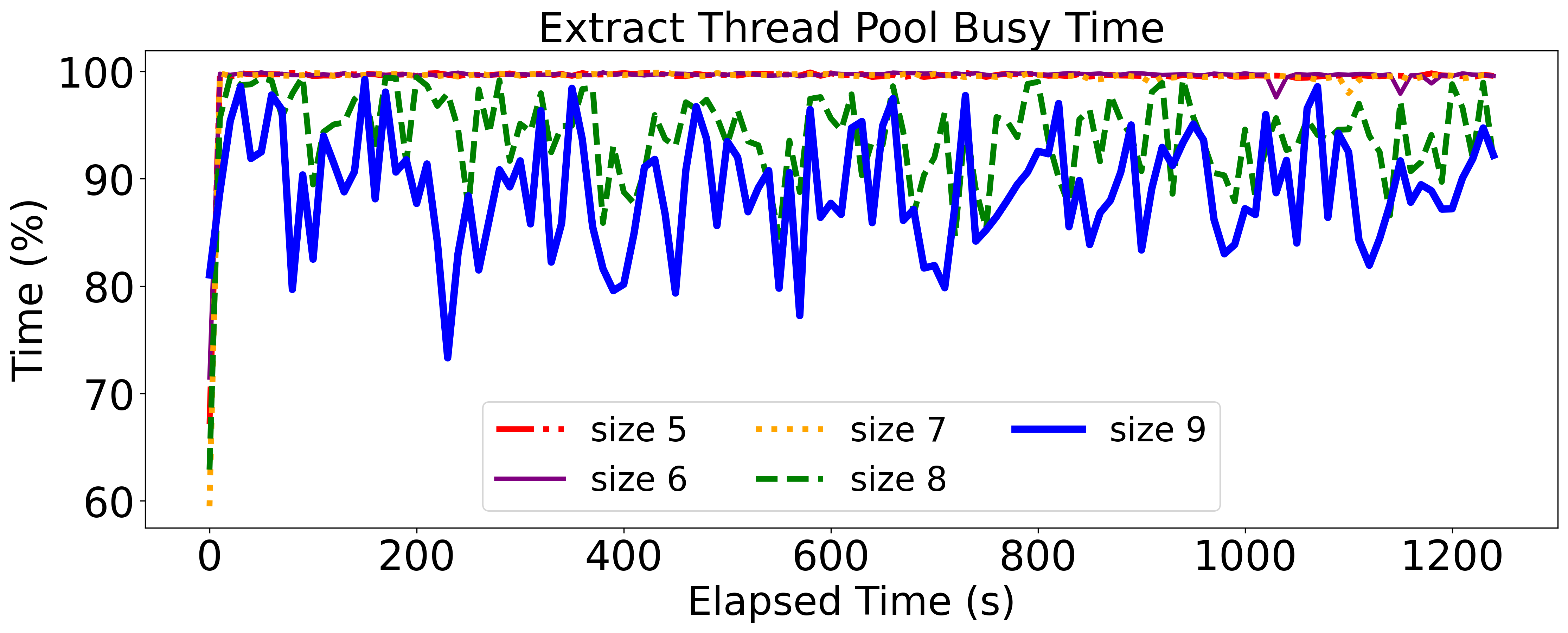}
  \caption{extract pool busy time.}
  \label{fig:sa_extract_threadpool_extract}
\end{subfigure}%
\begin{subfigure}{.49\textwidth}
  \centering
  \includegraphics[width=\linewidth]{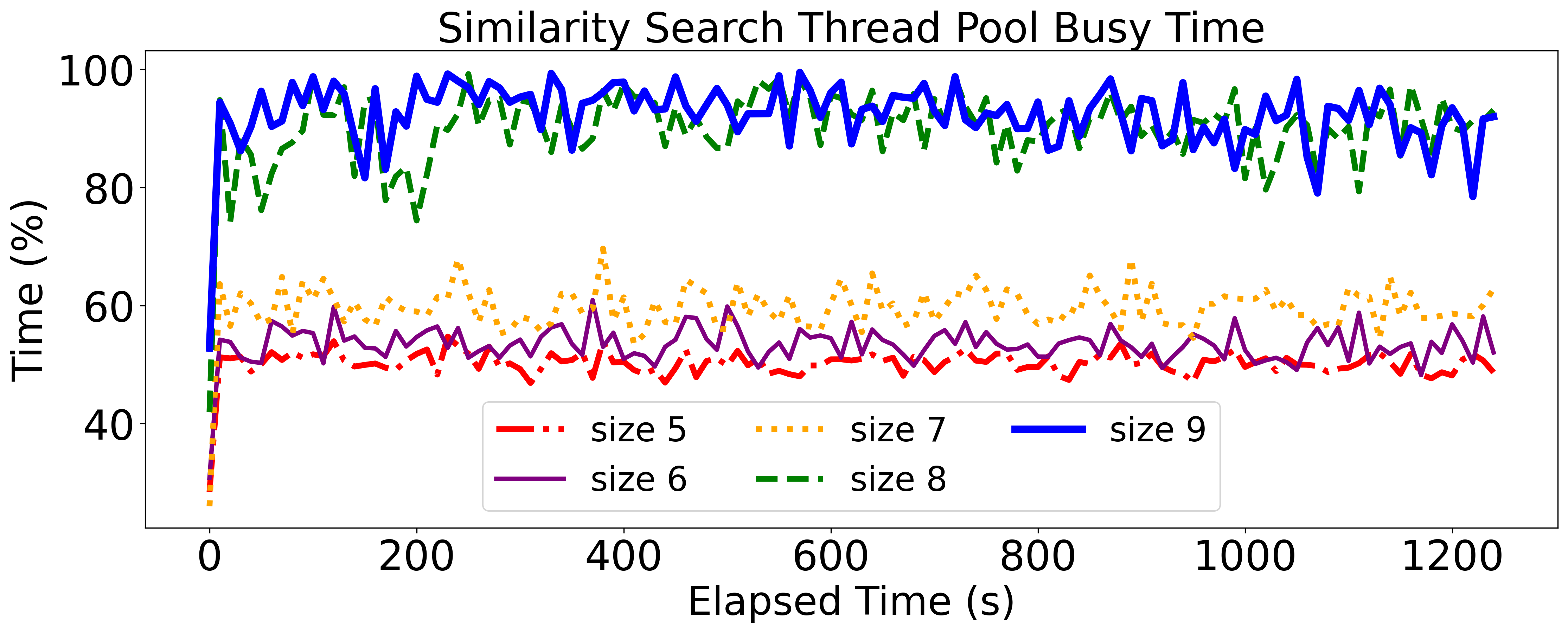}
  \caption{simsearch pool busy time.}
  \label{fig:sa_extract_threadpool_simsearch}
\end{subfigure}%

% \caption{Impact of extraction threads on: (a) user response time; (b) processing time; (c) CPU usage; (d) GPU memory usage; (e) system memory usage; (f) extract pool busy time; and (g) simsearch pool busy time.}
\caption{Impact of extract thread variability.}
\label{fig:sa_extract}
\end{figure*}

% \begin{figure}[t]
%   \centering
%   \includegraphics[width=0.99\linewidth]{fig_result_sa_simsearch.png}
%   \caption{Impact of similarity search threads on user response time.}
%   \label{fig:result_sa_simsearch}
% \end{figure}

\begin{figure*}[t]
\centering
\begin{subfigure}{.37\textwidth}
  \centering
  \includegraphics[width=\linewidth]{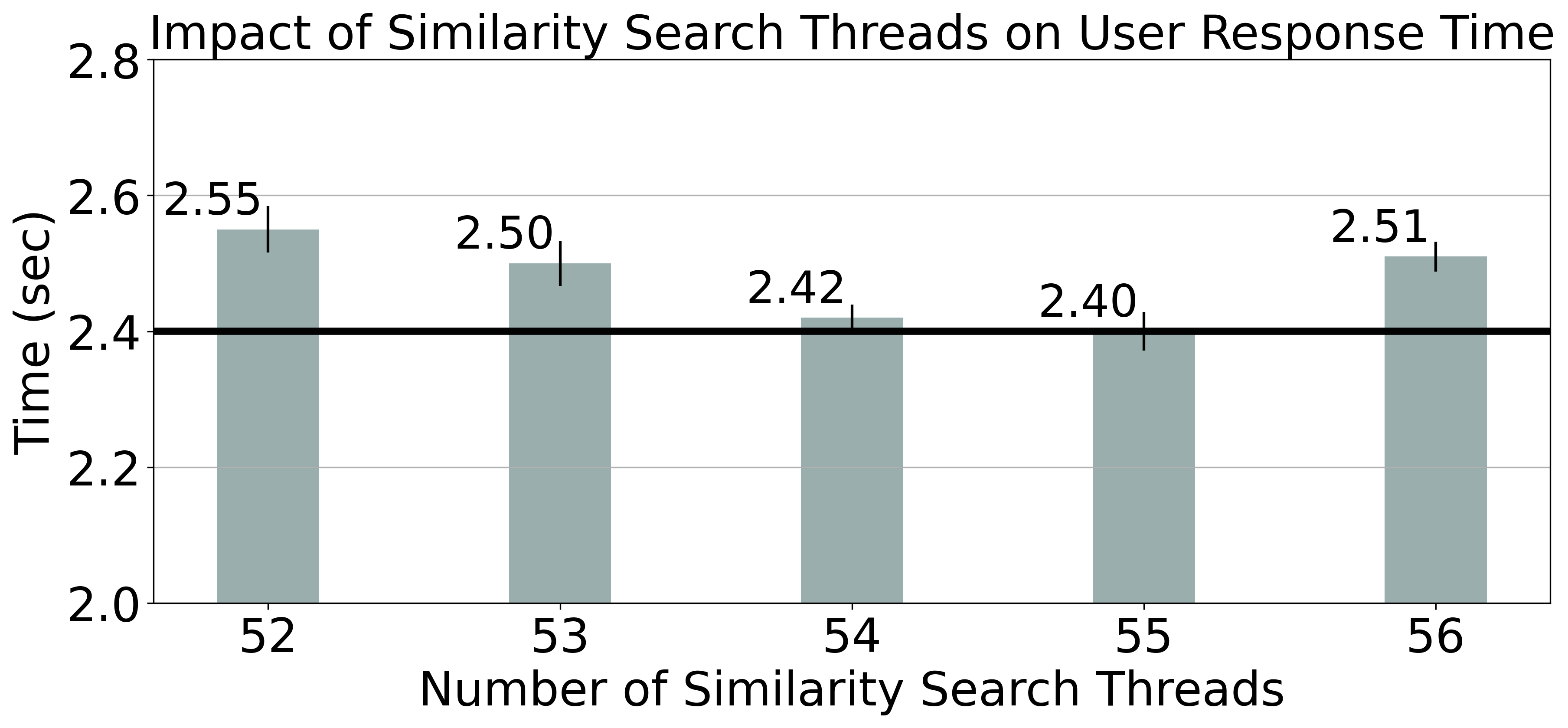}
  \caption{user response time.}
  \label{fig:result_sa_simsearch_rt}
\end{subfigure}%
\begin{subfigure}{.61\textwidth}
  \centering
  \includegraphics[width=\linewidth]{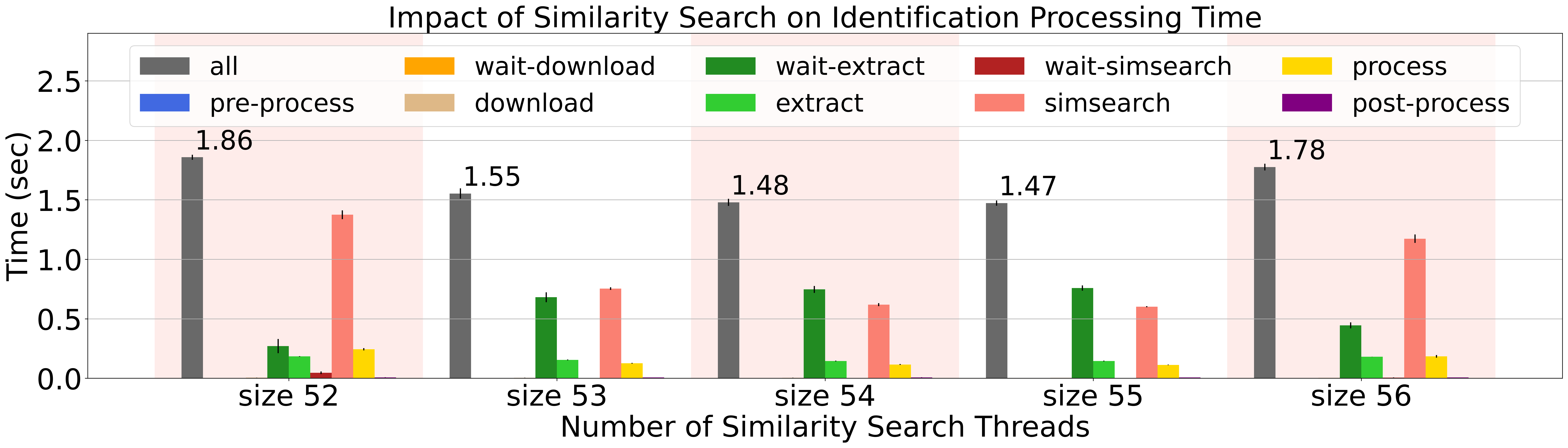}
  \caption{processing time.}
  \label{fig:result_sa_simsearch_pt}
\end{subfigure}

\begin{subfigure}{.49\textwidth}
  \centering
  \includegraphics[width=\linewidth]{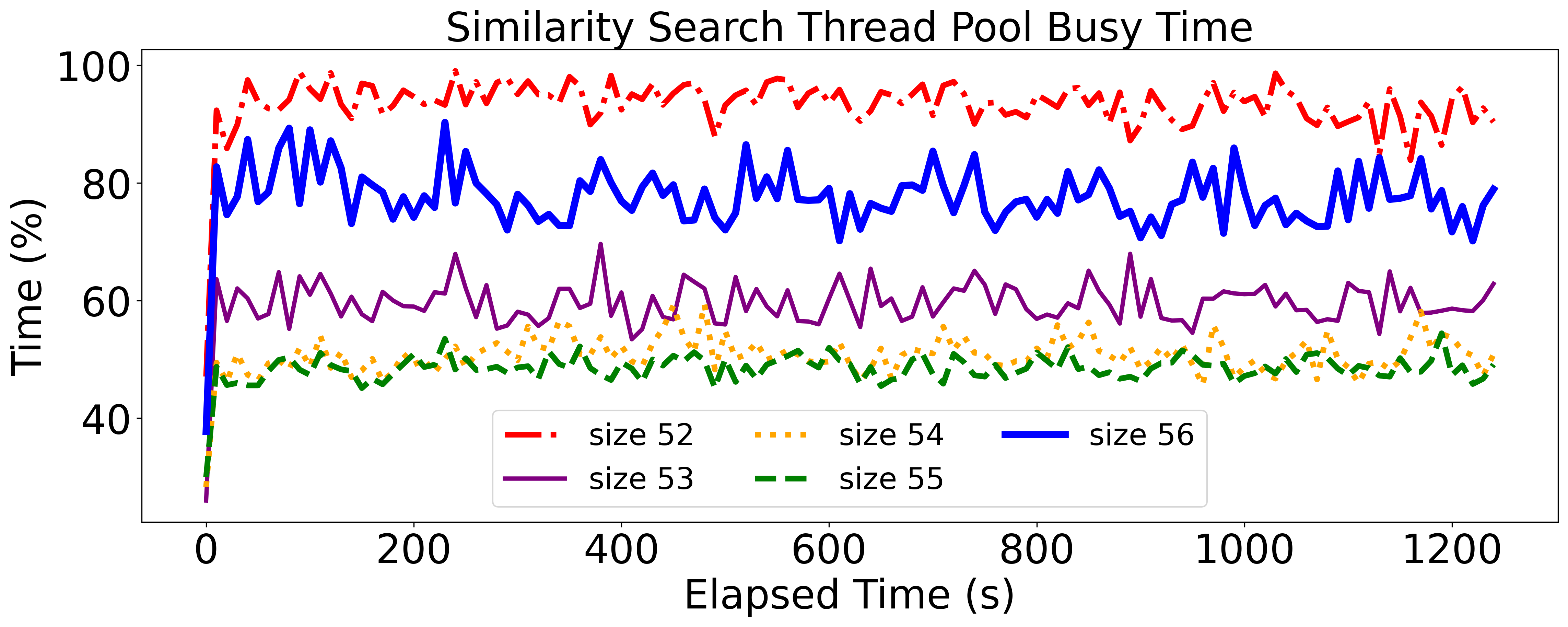}
  \caption{simsearch pool busy time.}
  \label{fig:sa_simsearch_threadpool_simsearch}
\end{subfigure}%
\begin{subfigure}{.49\textwidth}
  \centering
  \includegraphics[width=\linewidth]{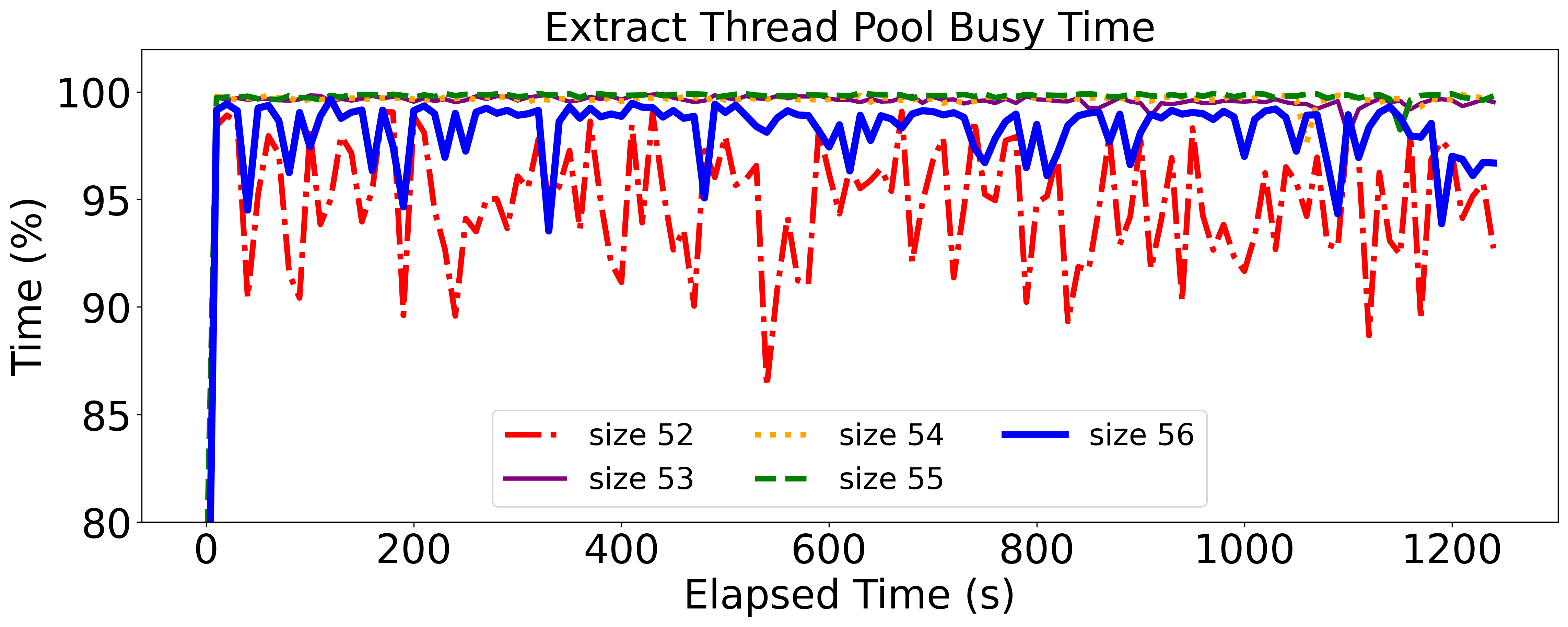}
  \caption{extract pool busy time.}
  \label{fig:sa_simsearch_threadpool_extract}
\end{subfigure}%

% \caption{Impact of similarity search threads on: (a) user response time; (b) processing time; (c) simsearch pool busy time; and (d) extract pool busy time.}
\caption{Impact of similarity search thread variability.}
\label{fig:sa_simsearch}
\end{figure*}

As presented in Figure~\ref{fig:result_baseline-vs-best-user-response-time}, we scale up the workloads as follows: 80, 120, and 140 simultaneous requests. As one may note, in Figure~\ref{fig:result_baseline-vs-best-user-response-time} the preliminary optimum configuration outperforms the baseline for all workloads. We highlight that, the difference between them varied as follows: 6.9\%, 2.2\%, and 6.7\% for 80, 120, and 140 simultaneous requests, respectively. %Furthermore, we observed that the baseline configuration crossed the 3-4 seconds constraint for a workload of 120 simultaneous requests, while the preliminary optimum one presented a \emph{user response time} of about 3.77 seconds in average.

The main observation is that the preliminary optimum configuration (found using our methodology) outperforms the baseline thanks to a better thread pool allocation that allows the Pl@ntNet system to serve simultaneously 35\% more requests (54 against 40) with a smaller user response time when compared to the baseline. We also highlight that, thanks to the transparent scaling feature provided by \textbf{E2C}\textit{lab}, one may easily scale up the workloads to analyze their impact on the application performance.

% Furthermore, the preliminary optimum configuration can deal with a workload of 120 simultaneous requests without crossing the 4 seconds constraint, in which case more than 60\% of the users abandon the transaction or even delete the mobile application. We also highlight that, thanks to the transparent scaling feature provided by \textbf{E2C}\textit{lab}, one may easily scale up the workloads to analyze their impact on the application performance.

% for the \emph{extract} and \emph{similarity search} tasks. 

\subsection{How do the \emph{Extraction} and \emph{Similarity Search} thread pool configurations impact the processing and user response times?}
\label{sec:sensitivity}

Since the \emph{extraction} and \emph{similarity search} tasks are the most time consuming compared to the remaining ones, we zoom our analysis on them in an attempt to improve even more the thread pool configuration and also to identify possible bottlenecks on the Pl@ntNet identification engine. The experiment aims to understand how variations in the preliminary optimum thread pool configuration of the \emph{extraction} and \emph{similarity search} tasks impact the user response time and the processing time of the identification tasks.

We apply \emph{Sensitivity Analysis} techniques to explore the impact of such variations. From the existing Sensitivity Analysis methods we decided to use \emph{One-at-a-time (OAT)}~\cite{hamby1995comparison}. OAT is a simple and common approach that consists in varying a single parameter at a time to identify the effect on the output. 

In our case, the parameters are \emph{extract} and \emph{simsearch} thread pool sizes. We vary the \emph{extract} pool size in $\pm2$ from the current size (7 threads), while the \emph{simsearch} in $\pm3$ (current size is 53 and for simplification, we do not present in Figure~\ref{fig:result_sa_simsearch_rt} the times for 50 and 51 since they are bigger than 52). These variations result in 10 new thread pool configurations to be evaluated. Therefore, we take advantage of \textbf{E2C}\textit{lab} to automatically run them in a reproducible way, following \textbf{E2C}\textit{lab}'s methodology.
% }

Figure~\ref{fig:sa_extract} shows the impact of extraction threads on: \emph{(a)} the user response time and \emph{(b)} the time to process each task. Furthermore, we also analyze their impact on resource usage, such as: (c) CPU usage (d) GPU memory, (e) system memory, (f) extract pool busy time, and (g) simsearch pool busy time.

In Figure~\ref{fig:result_sa_extract}, we observe that the preliminary optimum configuration with 7 extract threads does not produce the minimum user response time, since using 6 extract threads reduces it by 8.5\%. Decreasing to 5 threads or increasing it to 8 or 9 threads impacts negatively when compared to 6 threads. The explanation for this behaviour is given next.

Regarding the processing time (Figure~\ref{fig:result_sa_extract_pt}), as expected, the \emph{wait-extract} time reduces as we increase the number of \emph{extract} threads, while the \emph{simsearch} task time increases. This time increase in the \emph{simsearch} task can be explained by Figure~\ref{fig:sa_extract_cpu-usage}, since using 8 and 9 extract tasks results in a CPU usage of 100\% during the whole application execution, so as those tasks compete for processing resources, allocating more \emph{extract} threads impacts negatively on the \emph{simsearch} task time. As for the remaining sizes, they varied between 85\% and 100\%. This behaviour explains the results observed for the user response time presented in Figure~\ref{fig:result_sa_extract}. Furthermore, differently from the \emph{wait-extract} time, the \emph{extract} task time was not reduced when increasing the extract thread pool size.

\begin{table}[t]
\small
\centering
\caption{Comparison of the three Pl@ntNet configurations.}
\label{tbl:best-found-v2}
\begin{tabular}{llll}
\hline
\textbf{Thread pool}        & \textbf{baseline}        & \textbf{\begin{tabular}[c]{@{}l@{}}preliminary\\ optimum\end{tabular}}                             & \textbf{\begin{tabular}[c]{@{}l@{}}refined\\ optimum\end{tabular}}                          \\ \hline
\rowcolor[HTML]{E0EBEA} 
HTTP                        & 40                       & \cellcolor[HTML]{E0EBEA}54                      & \cellcolor[HTML]{E0EBEA}54                      \\
Download                    & 40                       & 54                                              & 54                                              \\
\rowcolor[HTML]{E0EBEA} 
Extract                     & 7                        & \cellcolor[HTML]{E0EBEA}7                       & \cellcolor[HTML]{E0EBEA}\textbf{6}              \\
Simsearch                   & 40                       & 53                                              & 53                                              \\ \hline
% \rowcolor[HTML]{E0EBEA} 
% \textbf{User resp. time} & \textbf{2.657 ($\pm 0.0914$)} & 
% \textbf{2.484 ($\pm 0.0912$)} & 
% \textbf{2.476($\pm 0.0826$)} \\ \hline

\textbf{\begin{tabular}[c]{@{}l@{}}User response \\time\end{tabular}} & \textbf{\begin{tabular}[c]{@{}l@{}}2.657 \\($\pm0.0914$)\end{tabular}} & \textbf{\begin{tabular}[c]{@{}l@{}}2.484 \\($\pm0.0912$)\end{tabular}} & \textbf{\begin{tabular}[c]{@{}l@{}}2.476 \\($\pm0.0826$)\end{tabular}} \\ \hline
\end{tabular}
\end{table}

By analyzing the impact on the GPU memory usage (Figure~\ref{fig:sa_extract_gpu-memory-usage}), we observe that it increases as we allocate more threads to the extract thread pool and it remains constant during the application execution. The GPU utilization for all thread pool sizes is between 35\% and 60\% most of the time, while the GPU power draw is between 50 Watts and 80 Watts. As the GPU memory usage, the system memory usage (Figure~\ref{fig:sa_extract_memory-usage}) of the Docker container running the Pl@ntNet Engine also increases with the extract thread pool size.

Lastly, the extract thread pool busy time (Figure~\ref{fig:sa_extract_threadpool_extract}) is 100\% during the whole application execution for thread pool sizes of 5, 6, and 7, and between 80\% and 100\%
for sizes of 8 and 9. This explains the higher and lower values, respectively, of the \emph{wait-extract} times observed in Figure~\ref{fig:result_sa_extract_pt}. For the similarity search (Figure~\ref{fig:sa_extract_threadpool_simsearch}), the thread pool busy time is between 80\% and 100\% for a size of 8 and 9. For the 5, 6, and 7 thread pool sizes it is 50\%, 55\%, and 60\% busy in average, respectively. This also explains the higher values of \emph{wait-simsearch} for sizes 8 and 9 compared to 5, 6, and 7 in Figure~\ref{fig:result_sa_extract_pt}.

Following our analysis, Figure~\ref{fig:sa_simsearch} shows the impact of the thread pool size for similarity search on: \emph{(a)} user response time and \emph{(b)} processing time. Besides, in Figure~\ref{fig:sa_simsearch_threadpool_simsearch} and Figure~\ref{fig:sa_simsearch_threadpool_extract} we show the thread pool busy time for the similarity search and extract thread pools, respectively. 

In Figure~\ref{fig:result_sa_simsearch_rt}, as one may note, the preliminary optimum configuration with 53 threads may be increased to 55 threads in order to reduce by about 4\% the user response time. Regarding the processing time (Figure~\ref{fig:result_sa_simsearch_pt}), the \emph{simsearch} task time confirms what was observed with the user response time, that is, adding more than 55 threads is not worth to decrease the execution time of the \emph{simsearch} task.

\begin{figure}[t]
  \centering
  \includegraphics[width=0.99\linewidth]{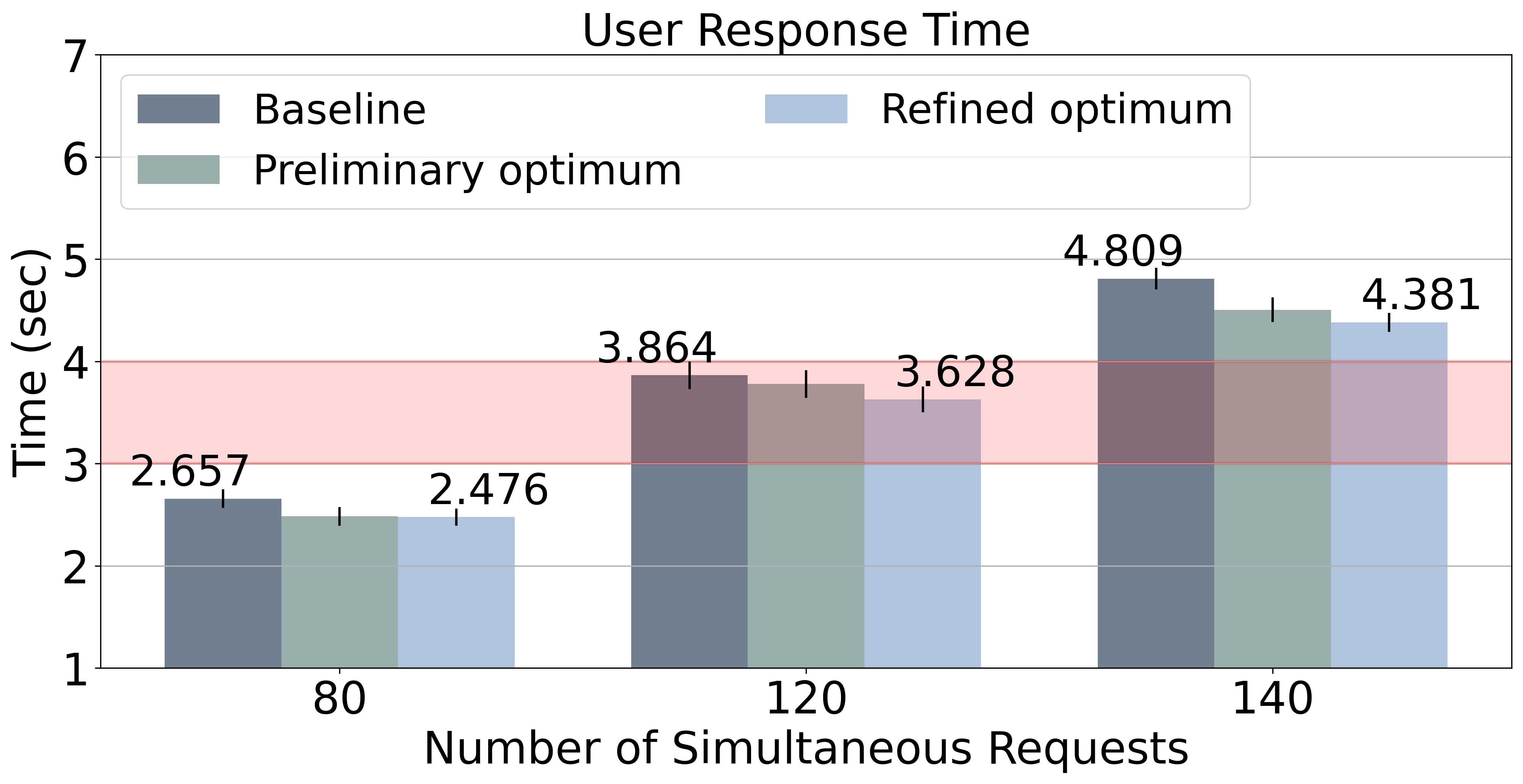}
  \caption{User response time: baseline vs optimums.}
  \label{fig:result_baseline-vs-best2-user-response-time}
\end{figure}

Figure~\ref{fig:sa_simsearch_threadpool_simsearch} shows the correlation of the similarity search pool busy time with the \emph{simsearch} task time observed in Figure~\ref{fig:result_sa_simsearch_pt} and explains its variation. Using 52 threads it is busy between 90\% and 100\%, while for 53 to 55 it is below 60\%, and increases to about 80\% with 56 threads. The impact of the similarity search thread pool variation on extract task (Figure~\ref{fig:result_sa_simsearch_pt}) can be explained by Figure~\ref{fig:sa_simsearch_threadpool_extract}. Lower times in \emph{wait-extract} for sizes 52 and 56 is due to a busy time between 90\% and 100\%. For sizes from 53 to 55, the busy time is 100\%.

Since we observed a lower user response time after analyzing the impact of variations of the \emph{extract} and \emph{simsearch} thread pool configurations on the user response time, we exploit this configuration (named \emph{\textbf{refined optimum}}) with all the previously defined workloads. As presented in Table~\ref{tbl:best-found-v2} and Figure~\ref{fig:result_baseline-vs-best2-user-response-time}, we observed even better results for all workloads.

Let us note that for all workloads the refined optimum presents the best results, outperforming both baseline and preliminary optimum. Compared with the baseline, the difference between configurations varied with the workloads as follows: from 6.9\% to 7.2\%; from 2.2\% to 6.3\%; and from 6.7\% to 9.8\% for 80, 120, and 140 simultaneous requests, respectively.

In summary, the analysis presented in this section backed by our optimisation methodology helped to understand how variations in the thread pool configuration of the Pl@ntNet engine impact on the processing times (user response time and identification processing steps) by correlating them with the resource usage. Furthermore, this analysis helps to improve the performance of the application by supporting 35\% more simultaneous users (54 against 40) and presenting a smaller \emph{user response time} for different workloads (80, 120, and 140 simultaneous requests) and 30\% less GPU memory utilization (7GB against 10GB), when compared to the baseline.

Let us also highlight that, despite our evaluations focusing on the Pl@ntNet as a use case, our methodology and its implementation in \textbf{E2C}\textit{lab} can be used to analyze other applications in the context of the Edge-to-Cloud Computing Continuum (more details in Section~\ref{sec:discussion-use-cases}).

\section{Discussion}
\label{sec:discussion}
The enhanced \textbf{E2C}\textit{lab} exhibits a series of features that make it a promising platform for future performance optimization of applications on the Edge-to-Cloud Continuum through reproducible experiments. We briefly discuss them here.

% \subsection{Reproducible experimental research}
\subsection{Reproducible application optimization}
\label{sec:discussion-reproducible}

Our optimization methodology %extends a previous one that 
is aligned with the Open Science~\cite{fecher2014open} goal to make scientific research processes more transparent and results more accessible. As presented in Section~\ref{sec:methodology}, it is implemented as an extension of the \textbf{E2C}\textit{lab} framework for reproducible experimentation across the Edge-to-Cloud Continuum.
%\textbf{E2C}\textit{lab} 
It provides guidelines to systematically define the whole optimization cycle, such as: \emph{(Phase-I)} defines the optimization problem and the application-related parameters to optimize; \emph{(Phase-II)} defines the sampling methods and the optimization techniques and hyperparameters; and \emph{(Phase-III)} provides access to the optimization results.

% Our methodology and its implementation are aligned with the Open Science~\cite{fecher2014open} goal to make scientific research processes more transparent and results more accessible. The \textbf{E2C}\textit{lab} framework provides a rigorous methodology that supports reproducibility. It provides guidelines to systematically define the whole experimental cycle, such as: the access to the experiment artifacts; the experimental environment; and the access to the experiment results.

The whole optimization cycle is defined through a configuration file (Listing~\ref{lis-udo}). This file was designed to be easy to use and to understand, and it can be easily adapted to different optimization problems (find out more in the documentation Web page~\cite{e2clab-doc-page}). At the end of each optimization cycle, \textbf{E2C}\textit{lab} provides an archive of the generated data. Such archive consists of data from \emph{Phases I and II}, needed to allow other researches to reproduce the research results. Regarding this work, the access to the experimental artifacts; definition of the experimental environment; and experimental results are publicly available at~\cite{e2clab-artifacts}. % may be found in the \emph{Artifact Description} and \emph{Artifact Evaluation} Appendix.

% Regarding this work, such archive may be found in our repository~\cite{e2clab-plantnet-repository}.

% The whole experimental cycle is defined through well structured configuration files. These files were designed to be easy to use and to understand, and to transparently scale the experimental environment to adapt it to different scenario configurations and allow users to analyze applications in different perspectives (find out more in the documentation Web page~\cite{e2clab-doc-page}).

% At the end of each experiment, \textbf{E2C}\textit{lab} provides an archive of the generated data. Such data refers to artifacts, experimental environment (\emph{e.g.} computational resources, network configuration, services configurations, and monitoring data), and experimental results. Regarding this work, such archive may be found in our repository~\cite{e2clab-plantnet-repository}.

\subsection{Scalable and parallel application optimization on large-scale testbeds}
\label{sec:discussion-scalable}
%Despite not well exploited in this work due to hardware constraints to run the Pl@ntNet application, \textbf{E2C}\textit{lab} supports the scalable and parallel optimization of applications on large-scale testbeds. Regarding \textbf{parallelism}, 

% \textcolor{blue}{
The proposed optimization methodology enables scalable (on large-scale testbeds), parallel (through asynchronous model training) and reproducible (by following a rigorous experimental methodology) application optimization. This approach speeds up the search of application parameters thanks to parallel and asynchronous application deployments on large-scale testbeds which helps to significantly reduce the application optimization time from days to hours compared to a sequential optimization approach.
% } 

% In our proposed optimization methodology, the evaluation of the application configuration in the search space may be \textbf{parallelized}, meaning that it would compute simultaneous objective function evaluations in parallel. To this purpose, our methodology implemented in \textbf{E2C}\textit{lab} takes advantage of the \emph{liar strategy}~\cite{chevalier2013fast} provided in Scikit-Optimize.

The parallel evaluation of the application configuration has the potential to scale to hundreds of machines in a large-scale testbed. Therefore, one may compute simultaneously 10, 20, or even more (depending on the testbed limits and the hardware requirements of the application) evaluations of the objective function to speed-up the computations. We plan to explore this potential in future work.

% \subsection{Supporting the Analysis of Other Application Use Cases}
\subsection{Optimizing other applications}
\label{sec:discussion-use-cases}
Our approach is generic: the optimization of other applications may be achieved by describing the application optimization problem in the \emph{optimization configuration file}. It allows one to define the optimization cycle and easily adapt it to different optimization application-specific problems.

% Our optimization methodology is not generic. The optimization of other applications may be achieved by defining application-specific optimization problems in the \emph{optimization configuration file}. As presented, such file allows one to define the optimization cycle and easily adapt it to different optimization problems related to each application.

Furthermore, users may easily apply our methodology to their applications thanks to the \emph{Services} abstraction provided by \textbf{E2C}\textit{lab}. \emph{Services} represent any system or a group of systems that provide a specific functionality or action in the scenario workflow. For instance, such services may refer to Flink, Spark or Kafka clusters, among others. 

In order to support their applications, users have to implement their \emph{User-Defined Services}. For this purpose, \textbf{E2C}\textit{lab} provides a \emph{Service} class in which users have to override a \emph{deploy} method to define the deployment logic of their services, such as: the distribution of services to the physical machines; and to install the required software to run these services. Next, \textbf{E2C}\textit{lab}'s \emph{Service} class provides a method to register the user services. Lastly, \textbf{E2C}\textit{lab} managers will be able to deploy each service on the testbed. Therefore, in the work described in this paper, we had to implement the Pl@ntNet service.

% \textcolor{blue}{
% Lastly, our optimization approach is based on Bayesian Optimization techniques. Bayesian Optimization is widely applied for global optimization of expensive-to-evaluate black-box functions. Applying Bayesian Optimization in real-life HPC simulations (\emph{e.g.}, chemistry, experimental particle physics, weather modeling, among others) has presented promising results, since it significantly reduces the number of simulations. Therefore, we believe that our proposed optimization methodology implemented in \textbf{E2C}\textit{lab} may also be useful for optimizing applications in the HPC domain.
% }

% \subsection{HW and SW generality}
% DONE: novelty in the implementation section
% DONE: table plantnet (cpu, gpu)
% DONE: justify and motivate optim algo
% DONE: clarify step to refined optimum: automatic SA
% 
% formulation optim problem

\section{Related Work}
\label{sec:related-work}

With the popularity of complex application workflows requiring hybrid execution infrastructures, the holistic analysis of such applications combining IoT Edge devices and Cloud/HPC systems has been a very active field of research in the last few years. 

Existing solutions focus on the simulation and emulation of \textbf{parts of the Edge-to-Cloud infrastructure}. EdgeCloudSim~\cite{sonmez2018edgecloudsim} is an environment for performance evaluation of Edge computing systems that provides simulation for Edge-based scenarios. Users may run experiments considering computational and networking resources. EmuFog~\cite{mayer2017emufog} is an extensible emulation framework for Fog-based scenarios that allows the emulation of real applications and workloads. However, they focus on the Edge and Fog layers separately, not on the Edge-to-Cloud Continuum as a whole.

In~\cite{kochovski2019architecture}, the authors proposed an approach for automated deployment (using Kubernetes~\cite{bernstein2014containers}) of Cloud applications in the Edge-to-Cloud Continuum. This approach explores methods for selection of the optimal infrastructure, satisfying \emph{QoS} requirements of Cloud applications. While, A3-E~\cite{baresi2019unified} provides a unified model for managing the life cycle of continuum applications (mobile, Edge, and Cloud resources). A3-E focuses on the placement of computation along the continuum based on the specific context and user requirements. However, both works fail on providing \textbf{configuration control of the parameters} of the application and of the underlying Edge-to-Cloud infrastructure; it is widely known and demonstrated that configuration strongly impacts performance. Thus, that support is essential for performing reproducible experiments.

In contrast, our optimization methodology integrates reproducibility by design, and its implementation within \textbf{E2C}\emph{lab} enables instrumentation of real-life applications on large-scale testbeds \textbf{across the entire Edge-to-Cloud Continuum}.

\section{Conclusions}
\label{sec:conclusions}

%The extended version of \textbf{E2C}\textit{lab} proposed in this work, provides a methodology for reproducible performance optimization of Edge-to-Cloud applications on large-scale testbeds. Thanks to its \emph{Layers} and \emph{Services} abstraction it can express several applications deployed on different environments, ranging from the Edge to the Cloud.

%As demonstrated in this work, such 
The optimization methodology proposed in this paper has proven useful for understanding and improving the performance of a real-life application used in production at large-scale. Thanks to the extension presented in this work, \textbf{E2C}\textit{lab} becomes, to the best of our knowledge, the first framework to support the complete deployment and analysis cycle of application workflows executed on the Computing Continuum, including deployment, configuration, monitoring, and gathering of results, and now performance optimization.

% PV: Another comment on Conclusion: don’t talk about future work, but give more details on the experimental validation. 

We have validated our proposed optimization methodology at large scale on 42 nodes of the \emph{Grid´5000} testbed. We have shown how it can be used to analyze and optimize the performance of the Pl@ntNet botanical application, used by more than 10 million users in 180 countries. 

% \textcolor{blue}{Over the years, Pl@ntNet has been used by more and more users and such strong growth in the number of users is leading to ever-increasing IT infrastructure costs. The thread pool allocation found using our methodology can process simultaneously 35\% more requests; presents a smaller user response time for different workloads; and consumes 30\% less GPU memory; when compared to the baseline. Serving 35\% more users per physical machine, in an highly accessed application like Pl@ntNet (\emph{i.e.,} with more than 10M users), may significantly reduce the annual overall costs involved to maintain the Pl@ntNet infrastructure and keep the application free and accessible to all.} Despite our focus on Pl@ntNet, the methodology can be generalized to other applications in the Edge-to-Cloud Continuum.

The thread pool allocation found using our methodology increases the number of simultaneous requests processed in parallel by 35\% compared to the baseline; it reduces user response time for different workloads; and consumes 30\% less GPU memory. Despite our focus on Pl@ntNet, the methodology can be generalized to other applications in the Edge-to-Cloud Continuum.

%\textcolor{blue}{
%\textbf{E2C}\textit{lab} is aligned with the Open Science goal to make scientific research processes more transparent and results more accessible. Therefore, for reproducibility purposes, all the experimental results produced in this paper are publicly available and more details may be found in the \emph{Artifact Description} and \emph{Artifact Evaluation} Appendix.}

% In future work, we plan to focus on three directions: \emph{(1)} adding support for other parallel ML-based optimization techniques and explore their scalability in large-scale testbeds; \emph{(2)} enabling \emph{built-in} support for other large-scale experimental testbeds, besides Grid'5000, such as Chameleon (currently, users can adapt \textbf{E2C}\textit{lab} to their testbed of convenience by means of EnOSlib~\cite{cherrueau2018enosstack}); and \emph{(3)} enhancing reproducibility aspects by exploring techniques for provenance collection of data generated from running experiments, the goal is to provide additional context that more accurately explains the experiment execution and results.

\section*{Acknowledgments}
This work was funded by Inria through the HPC-BigData Inria Challenge (IPL) and by French ANR OverFlow project (ANR-15- CE25-0003). Experiments presented in this paper were carried out using the Grid'5000 testbed, supported by a scientific interest group hosted by Inria and including CNRS, RENATER and several Universities as well as other organizations. We also would like to thank Romain Egele, Jaehoon Koo, Prasanna Balaprakash, and Orcun Yildiz from Argonne National Laboratory for their support.

% \begin{itemize}
% \item funding agency
% \end{itemize}

\balance
\bibliographystyle{IEEEtran}
\bibliography{references}

\end{document}